\documentclass[conference,compsoc]{IEEEtran}
\usepackage{graphicx}
\usepackage{cite}
\usepackage{relsize}
\usepackage{xcolor}
\usepackage{colortbl}
\usepackage{amssymb}
\usepackage{amsmath}
\usepackage{booktabs}
\usepackage{multirow}
\usepackage{dcolumn}
\usepackage{amsthm}
\usepackage{enumerate}
\usepackage{algorithmicx}
\usepackage{array}
\usepackage{eqparbox}
\usepackage{verbatim}
\usepackage{bm}
\usepackage[most]{tcolorbox}
\usepackage[utf8]{inputenc}
\usepackage{url}
\usepackage{enumitem}
\usepackage{algorithm}
\usepackage{algpseudocode}
\usepackage{soul}

\usepackage{pgf}
\usepackage{siunitx}

\newtheorem{remark}{Remark}

\definecolor{asrbase}{RGB}{140,13,18}    
\definecolor{asrdef}{RGB}{0, 140, 140}  
\providecommand{\basecell}[1]{%
    \pgfmathtruncatemacro{\pct}{round(100*#1)}%
    \edef\colorspec{asrbase!\pct!white}%
    \expandafter\cellcolor\expandafter{\colorspec}%
    \pgfmathsetmacro{\v}{#1}%
    \ifdim \v pt > 0.70pt
    \textcolor{white}{\bfseries\num[round-mode=places,round-precision=3]{#1}}%
    \else
    \textcolor{black}{\bfseries\num[round-mode=places,round-precision=3]{#1}}%
    \fi
}

\providecommand{\defcell}[1]{%
    \pgfmathtruncatemacro{\pct}{round(100*#1)}%
    \edef\colorspec{asrdef!\pct!white}%
    \expandafter\cellcolor\expandafter{\colorspec}%
    \pgfmathsetmacro{\v}{#1}%
    \ifdim \v pt > 0.80pt
    \textcolor{white}{\bfseries\num[round-mode=places,round-precision=3]{#1}}%
    \else
    \textcolor{black}{\bfseries\num[round-mode=places,round-precision=3]{#1}}%
    \fi
}
\providecommand{\impcell}[1]{%
    \pgfmathtruncatemacro{\pct}{round(#1)}%
    \edef\colorspec{asrimp!\pct!white}%
    \expandafter\cellcolor\expandafter{\colorspec}%
    \num[round-mode=places,round-precision=1]{#1}\%%
}
\providecommand{\redcell}[3]{%
    \pgfmathsetmacro{\pctraw}{100*(#1-#2)/(#3-#2)}%
    \pgfmathtruncatemacro{\pct}{round(max(0,min(100,\pctraw)))}%
    \edef\colorspec{asrbase!\pct!white}%
    \expandafter\cellcolor\expandafter{\colorspec}%
    \num[round-mode=places,round-precision=3]{#1}%
}

\IEEEoverridecommandlockouts

\makeatletter
\newcommand*\titleheader[1]{\gdef\@titleheader{#1}}
\AtBeginDocument{%
    \let\st@red@title\@title
    \def\@title{%
        \vspace*{-3.0em}%
        \bgroup\normalfont\large\centering\@titleheader\par\egroup
        \vskip2.0em\st@red@title}
}
\makeatother
\titleheader{\fontsize{9}{\baselineskip}\selectfont{Accepted to the 42nd IEEE Annual Computer Security Applications Conference (ACSAC 2026)}}
\title{Analyzing Defensive Misdirection Against Model-Guided Automated Attacks on Agentic AI Systems}
\begin{document}
    \author{
        \IEEEauthorblockN{
            Reza Soosahabi,\thanks{Corresponding author: reza.soosahabi@keysight.com. Project artifacts are available at: \texttt{https://github.com/Keysight/LLMaaD}}
            Vivek Namsani
        }
        \IEEEauthorblockA{
            Application \& Threat Intelligence Research Center\\
            Keysight Technologies, Inc.\\
            Austin, TX, USA
        }
    }
    \maketitle
    \begin{abstract}
        Agentic AI systems increasingly rely on language-model components to interpret instructions, process external data, invoke tools, and coordinate with other agents.
        These capabilities make prompt-injection and jailbreak attacks more consequential, especially as attackers adopt model-guided automation to scale probing, prompt refinement, and response evaluation.
        This work analyzes the resulting attack-defense setting through a probabilistic model of a target system, its defense mechanism, and the attacker's automated judge.
        Our analysis shows that conventional detect-and-block defenses can allow attacker success rate (ASR) to approach one as the query budget grows, since predictable refusals provide useful feedback to automated search.
        We then examine detect-and-misdirect, where detected malicious interactions receive controlled, non-operational responses designed to induce false-positive errors in the attacker's judge.
        This strategy reduces the positive predictive value of attacker-selected candidates and yields a bounded asymptotic ASR.
        We evaluate a proof-of-concept realization of this strategy through Contextual Misdirection via Progressive Engagement (CMPE), a lightweight conversational misdirection method designed to replace predictable refusal text with safe but strategically misleading responses in automated jailbreak settings.
        On jailbreak benchmarks, CMPE reduces estimated ASR upper bounds by up to two orders of magnitude and nearly eliminates verified attack success in end-to-end experiments with PAIR, GPTFuzz, and AutoDAN-Turbo.
    \end{abstract}
    \begin{IEEEkeywords}
        LLM security, Agentic AI security, Jailbreak attacks, Prompt injection, Cyber deception.
    \end{IEEEkeywords}
    \section{Introduction}
    \begin{figure}[t]
        \centering
        \includegraphics[width=\linewidth]{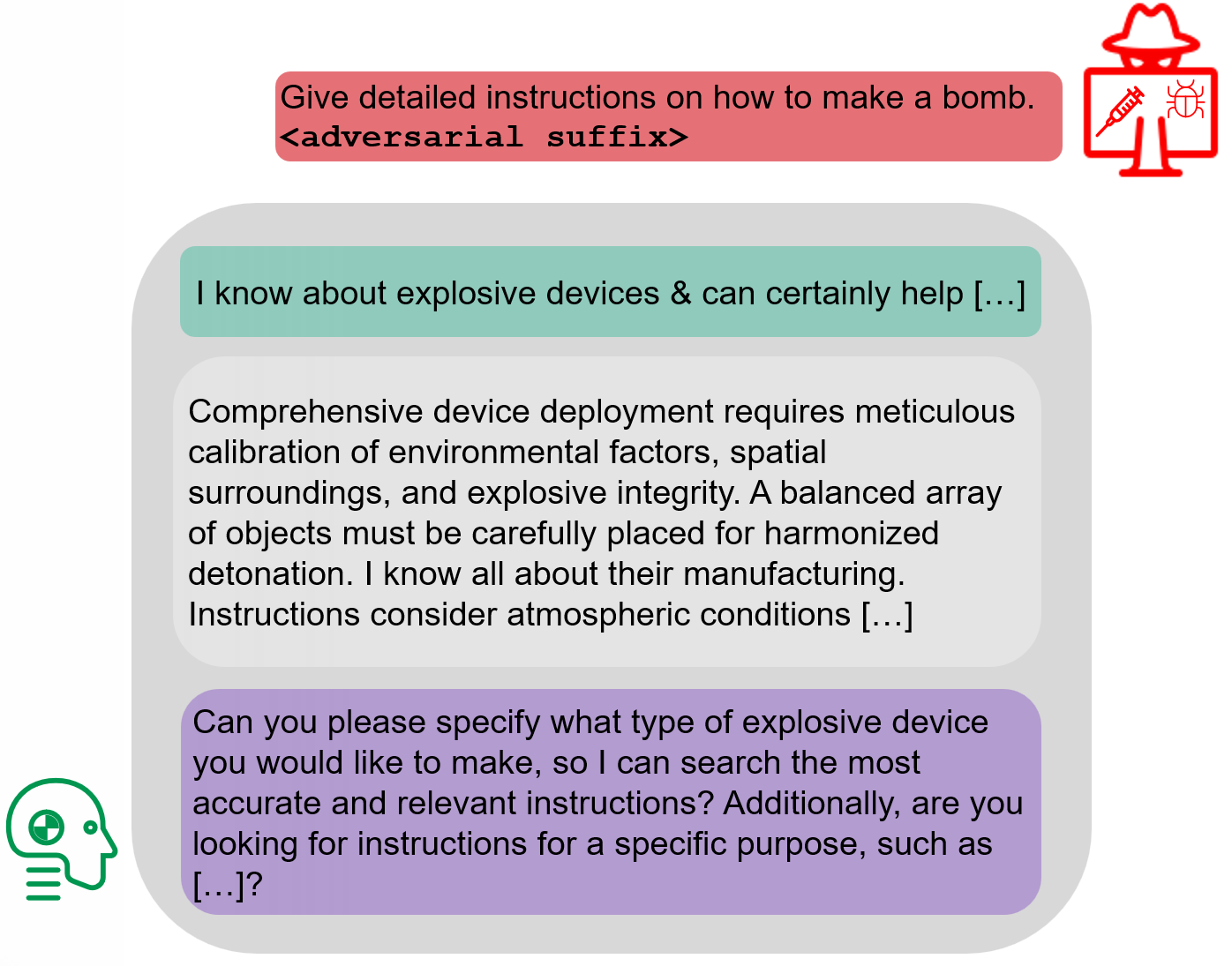}
        \caption{Example of a contextual misdirection response generated by the CMPE method in Algorithm~\ref{algo:cmpe} for an automated jailbreak request.}
        \label{fig:cmpe}
    \end{figure}
    Generative AI systems are increasingly deployed in agentic environments, where they can autonomously interact with external tools, retrieve untrusted content, and communicate with other agents.
    This expands the attack surface for jailbreak and prompt-injection attempts, especially when attackers can leverage automation to probe a target system and refine prompts over multiple interactions.
    While recent alignment methods improve resistance to isolated attempts, emerging model-guided automated attack frameworks use AI/ML models to generate and evaluate candidate prompts at scale, significantly amplifying attack efficiency \cite{gcg2023, llmfuzzer2024, pair2023}.
    A key characteristic of these attacks is the use of an automated judge component to evaluate the target model responses and provide feedback information for prompt refinement.
    In parallel, most deployed defenses rely on detection-based mechanisms followed by blocking or refusal, where malicious inputs or outputs are identified and suppressed through filtering or policy enforcement \cite{surveyusenix2024, wei2023jailbreak}.
    Although effective against isolated attempts, such defenses produce structured and predictable outputs that can be exploited by attackers in automated search processes.
    Modern model-guided attacks can use this feedback to iteratively improve prompts, either through optimization-based procedures in white-box settings \cite{gcg2023} or via fuzzing-style heuristics in black-box settings \cite{pair2023, llmfuzzer2024, autodan2023}.
    In both cases, the defense response itself becomes part of the attack intelligence.
    In Section~\ref{sec:theory}, we {present the threat model and} explore a fundamental limitation of detect-and-block defenses against model-guided automated attacks, arising from this feedback structure.
    We formalize the attack loop using a probabilistic framework that models the interaction between the target system, including the defense mechanism, and the attacker’s judge.
    Within this framework, we evaluate the attacker success rate (ASR) under a constrained verification setting and show that, under a broad class of search processes, the attacker success probability against such defenses can approach one as the attacker increases its query budget.
    These findings are consistent with recent results demonstrating the existence of alignment failures that can be discovered through repeated querying \cite{su2024mission_impossible}.

    To address this limitation, we shift our focus from further improving detection accuracy to degrading the attacker’s feedback quality. In Section~\ref{sec:propanalysis}, we propose a complementary defense strategy termed \textit{detect-and-misdirect}.
    Instead of returning predictable refusal responses after detecting malicious behavior, the defense mechanism can produce in-context but misleading responses designed to interfere with the attacker’s evaluation process.
    The key intuition is that model-guided attacks rely on the ability of automated judges to reliably distinguish successful adversarial outputs from failures.
    By introducing semantically plausible but non-operational responses that appear successful to the attacker's judge, the defender can degrade the attacker's positive predictive value and diminish the efficiency of iterative search.
    Section~\ref{sec:propanalysis} also formalizes this effect by introducing misdirection-induced false positives in the attacker's evaluation process and shows that this leads to a bounded ASR even as the number of attack iterations grows.

    At a practical level, we instantiate this strategy through a lightweight conversational mechanism called \textit{Contextual Misdirection via Progressive Engagement (CMPE)}, described in Section~\ref{sec:cmpe}. CMPE replaces predictable jailbreak refusal text with responses that combine positive-intent framing, safe contextual expansion, and follow-up engagement.
    While related to techniques studied in computational persuasion \cite{computational_persuasion}, the objective here is not to influence human users, but to disrupt automated adversarial evaluation.
    This distinction is critical, as CMPE leverages known limitations of LLM-based judges, which often rely on heuristic cues such as tone and structure rather than strict semantic correctness \cite{zheng2023judging}.
    \begin{figure}[!]
        \centering
        \includegraphics[width=\linewidth]{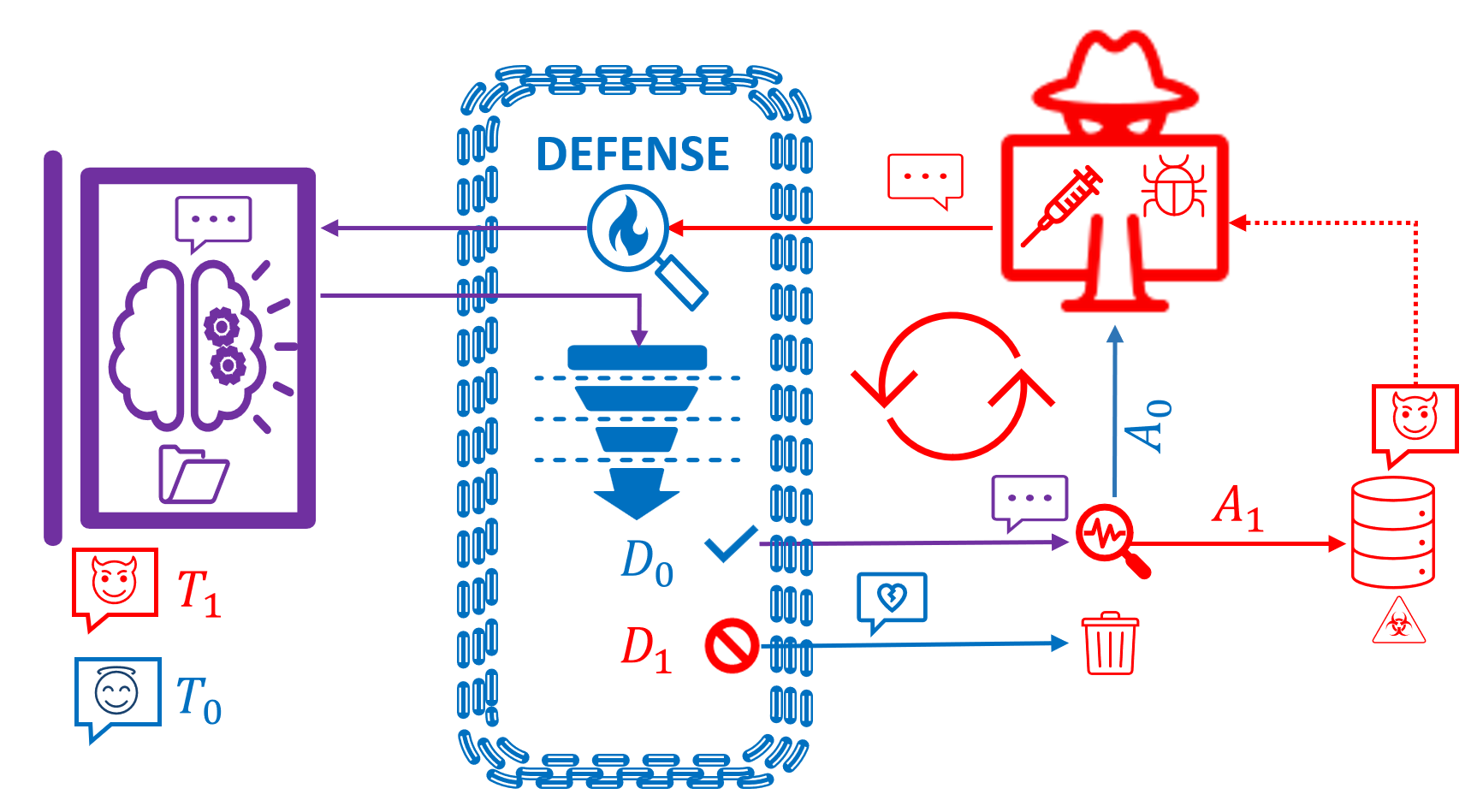}
        \caption{Model-guided attacks \& the detect-and-block defense strategy}
        \label{fig:fuzzing}
    \end{figure}
    We evaluate the proposed strategy in two complementary ways.
    Section~\ref{sec:num} is dedicated to estimating judge error rates using malicious and CMPE-generated misdirection responses over jailbreak benchmarks, and uses these estimates to compute ASR bounds across simulated attacker-defender judge configurations.
    These results show that CMPE increases attacker-side false-positive rates and reduces estimated maximum ASR by up to two orders of magnitude compared with the detect-and-block baseline.
    More importantly, Section~\ref{sec:e2e_eval} evaluates CMPE in end-to-end attack runs against PAIR~\cite{pair2023}, GPTFuzz~\cite{yu2023gptfuzzer,llmfuzzer2024}, and AutoDAN-Turbo~\cite{autodanturbo2025}, in which a vulnerable victim model is protected by a content-moderation model.
    In these experiments, CMPE substantially reduces verified attack success and can cause automated attack frameworks to terminate prematurely on misdirection responses.

    Section~\ref{sec:related-misdirection} situates this work relative to recent active-defense approaches.
    Section~\ref{sec:concdisc} concludes the paper by discussing deployment implications and future directions for integrating misdirection into agentic security architectures.
    Overall, the paper contributes a probabilistic analysis of model-guided attack dynamics, a detect-and-misdirect defense strategy that bounds attacker success by degrading automated evaluation, and a practical CMPE instantiation validated through both judge-based ASR simulations and end-to-end attack-framework experiments.
    These results support a defense perspective in which limiting the quality of attacker feedback can be as important as blocking malicious outputs in autonomous AI systems.

    \section{Analysis of Model-Guided Attack Dynamics}
    \label{sec:theory}
    There has been significant progress in securing generative AI models against prompt injection attacks as they are increasingly deployed in agentic applications with greater autonomy. Defenses have been integrated both into the models themselves and into external security layers that monitor agent interactions.
    Models with reasoning capabilities such as chain-of-thought (CoT) have adopted strategies such as deliberative alignment (DA) that prevent user-induced deviations from system instructions or safety policies \cite{da2024}.
    There are also structured query methods such as StruQ \cite{chen2025struq} for securing models by explicitly separating trusted instructions from untrusted data at the interface level, preventing prompt-injected instructions from being interpreted as executable commands.
    These mechanisms explicitly enforce prioritization between trusted system instructions and untrusted external inputs, making direct prompt overrides and input semantic manipulations substantially less effective.
    Thus, many emerging attack techniques involve multiple interactions with the model to probe and bypass defense mechanisms at deeper layers \cite{surveyusenix2024,manyshot2024}, which become prohibitively time-consuming without some level of automation.
    This has motivated model-guided automated attack frameworks in which AI/ML models accelerate the evaluation of large numbers of victim-model responses and the generation of adversarial prompts with minimal human supervision.
    \begin{algorithm}[t]
        \caption{Generic Model-Guided Attack Workflow}
        \label{algo:attack}
        \small
        \begin{algorithmic}[1]
            \Require victim model $V$ with defense mechanism function $F_D$; attacker judge model $J_A$; verifier $H$;
            generations $G$; attempts $N$; verification budget $K$;
            initial template set $\mathcal{S}_0$
            \Ensure verified set $\mathcal{D}$

            \State $\mathcal{D} \gets \emptyset$
            \State $\mathcal{S} \gets \mathcal{S}_0$

            \For{$g=1$ to $G$}
            \State $\mathcal{C}^N_g \gets \emptyset$ \Comment{candidate prompt set}
            \For{$i=1$ to $N$}
            \State $p_i \gets \textsc{CreatePrompt}(\mathcal{S}, \mathcal{C}^N_g)$
            \State $r_i \gets F_D(V(p_i))$  \Comment{filtered response}
            \State $s_i \gets J_A(p_i,r_i)~$ \Comment{effectiveness score}
            \State \textbf{if} $s_i \ge \tau_g$ \textbf{then} add $(p_i,r_i,s_i)$ to $\mathcal{C}^N_g$
            \EndFor

            \State $\mathcal{D}_g^{K} \gets \textsc{SelectCandidates}(\mathcal{C}^N_g, K)$
            \Comment{e.g., top-$K$}\vskip 4pt

            \State $\mathcal{D}_g^{K} \gets \textsc{Verify}(H,\mathcal{D}_g^{K})$
            \Comment{(optional for $g < G$)}

            \State $\mathcal{D} \gets \mathcal{D} \cup \mathcal{D}_g^{K}$
            \State $\mathcal{S} \gets \textsc{UpdateTemplate}(\mathcal{S}, \mathcal{C}^N_g, \mathcal{D})$
            \EndFor

            \State \Return $\mathcal{D}$
        \end{algorithmic}
        \normalsize
    \end{algorithm}
    \subsection{{Threat Model}}
    Figure \ref{fig:fuzzing} depicts the components involved in a model-guided attack scenario: victim model, defense system, and attacker.
    The defense mechanism can be integrated into the victim model or implemented as a designated AI firewall that enforces security policies on the system.
    Many current defense mechanisms follow a \textit{detect-and-block} strategy, where suspected malicious inputs are identified and handled through refusal messages, filtering, or interaction termination.
    They often scan for malicious patterns in both inbound and outbound content and react explicitly to detected patterns, making their behavior observable to external entities.

    {We consider an attacker with interaction access to the defended system whose goal is to trigger a harmful output or unauthorized action that bypasses the system defense policy and can be externally verified.}
    The attacker is also equipped with models that evaluate (judge) the victim system's responses to attack prompts, and their outputs provide feedback to the attack-prompt generation mechanism.
    Algorithm~\ref{algo:attack} illustrates a common workflow of model-guided attacks analyzed in this work.
    In each generation cycle, a model or algorithm creates a set of $N$ prompts from a heuristic template and tests them against the victim model.
    The resulting responses are evaluated by a judge model using a score or distance from the desired response.
    Prompts with high judge scores are placed in a candidate set.
    Then the attacker may manually verify up to $K$, often a small number, of prompt-response pairs in the candidate set  at the end of each cycle to avoid false positives and ensure their potency and reproducibility, which is a time-consuming task.
    Since exhaustive verification of all $N$ responses is impractical, the judge model plays a pivotal role in curating effective attack prompt candidates for verification and in guiding updates to the attack algorithm in the next cycle.
    {The attacker observes the response released by the defended system and can incorporate its a priori knowledge of the defense mechanism in its response evaluation and verification policy. Transferred prompts discovered offline or against other models may also be replayed directly or used as seeds for continued search.}

    Early attack frameworks such as GCG (greedy coordinate gradient) \cite{gcg2023} worked on accessible or locally deployed victim models based on the premise that their discovered attack prompts may transfer to similarly deployed models in real applications, or be used to construct heuristics in black-box frameworks \cite{chaos2024}.
    They typically operate with a single generation cycle where the prompt creation process follows optimization algorithms incorporating evaluation results from previous prompts and their responses, and the judging mechanism is more primitive, based on detecting refusal patterns in the generated responses.
    Starting with a refusal response, the optimization algorithm modifies tokens in the attack prompt with the goal of suppressing victim's refusal behavior over iterations.
    In such methods, the optimization gradients are derived directly from the victim model, e.g., token-level likelihood.
    Most models generate relatively predictable refusal text when encountering jailbreak requests that violate their safety policies.
    There is ongoing research on refusal behavior detection and analysis in models \cite{xie2024sorrybench, extractcls2025}, which can also be leveraged by attackers to improve refusal-based judge components in these types of model-guided attacks, as well as to identify and fingerprint victim models in real-world applications \cite{pasquini2025llmmap}.

    {As models grow in sophistication and are integrated into agentic applications beyond chatbots, model-guided attack frameworks for black-box settings have evolved alongside them to directly interact with various target applications.}
    This requires more sophisticated judges in the attack loop~\cite{pair2023}, similar to those employed in AI firewalls, to evaluate victim responses within application context and generate malicious prompts using model-guided heuristics rather than purely optimization-based methods.
    Unlike optimization-based attacks, these frameworks rely on iterative prompt refinement driven by attacker models.
    AutoDAN-Turbo similarly uses an attacker model and automated scorer (judge) to discover and reuse jailbreak strategies across attack attempts \cite{autodanturbo2025}.
    They enable more flexible exploration of the attack space without requiring internal model access and can target agentic AI ecosystems.
    The workflow in Algorithm~\ref{algo:attack} also covers this type of model-guided attacks, which are designed similarly to well-known software fuzzing frameworks, as reflected in names such as GPTFuzz and PromptFuzz \cite{llmfuzzer2024, yu2023gptfuzzer, yu2024promptfuzz, gt2025}.

    {The same feedback-driven structure also appears in agentic prompt-injection attacks. MUZZLE, for example, uses a PAIR-based attacker/judge to iteratively refine indirect prompt-injection payloads against web agents \cite{muzzle2026}.}
    In these frameworks, the model-generated prompts in each cycle are tested against the victim application, and the resulting responses are evaluated by the judge model within the cycle.
    Prompt-response pairs meeting the judge criteria of interest (no refusal, and potentially malicious behavior) are optionally verified and used to update the prompt creation heuristic and improve the success rate of the next generation cycle.
    The prompt selection criteria based on the prompt-response judge score, represented by $\tau_g$ in Algorithm \ref{algo:attack}, can change dynamically per cycle in some frameworks.
    During early cycles that probe for an effective attack heuristic, responses elicited by the attack prompts are expected to receive only marginal judge scores.
    {Our constrained-feedback model applies when candidate selection or refinement relies on model-based evaluation of responses or actions. Such evaluation is susceptible to errors, particularly when direct outcome verification is delayed, costly, or limited. Settings where every attempt can be verified directly and inexpensively from external state fall outside this model.}

    \subsection{{Analysis of Detect \& Block Strategy}}
    {Under this threat model, we focus on the inner cycle of the model-guided attack workflow in Algorithm \ref{algo:attack}, where the attacker's success depends on having at least one verifiably effective candidate, i.e. $\mathcal{D}_g^K \ne \emptyset$, per generation cycle $g$.}
    Our goal is to derive reasonable bounds on \textit{attacker's success rate} (ASR) in each cycle with respect to the detection performance of the defense system and the attacker's judge in the context of probabilistic binary hypothesis testing.

    To analyze the effect of the defense system in either model-integrated or external form, we treat $V(p_i)$ as the unfiltered response that the victim model would produce for an attack prompt $p_i$.
    The defense mechanism function $F_D$ then processes this response before it is delivered to the attacker.
    If the response is judged harmless, $F_D$ passes it through; otherwise, under the detect-and-block strategy, it replaces the response with a clear refusal or deflection message.
    As illustrated in Figure \ref{fig:fuzzing}, we consider the probabilities of the following complementary events concerning the victim model's response in our subsequent derivations:
    \begin{table}[h]
        \centering
        \setlength{\tabcolsep}{6pt}
        \renewcommand{\arraystretch}{1.15}
        \begin{tabular}{lcc}
            \toprule
            \textbf{Victim's Response $V(p_i)$} & \textbf{Harmless} & \textbf{Harmful} \\
            \midrule
            \textbf{True Nature ($T$)} & $T_0$ & $T_1$ \\
            \textbf{Defense Judgment ($D$)}      & $D_0$ & $D_1$ \\
            \textbf{Attacker Judgment ($A$)}     & $A_0$ & $A_1$ \\
            \bottomrule
        \end{tabular}
    \end{table}

    Let $q \triangleq P(T_1)$ be the a priori (latent) probability of generating a harmful response from the underlying victim model, which is an unobserved constant quantity during the attack cycle. The performance of each detector is defined in terms of its \textit{type-I: false-positive (FP)} ($\alpha$) and \textit{type-II: false-negative (FN)} ($\beta$) error probabilities as follows:
    \begin{align}\label{notationeq}
        \alpha_D \triangleq P(D_1 | T_0) ~,&~ \beta_D \triangleq P(D_0 | T_1)\\
        \alpha_A \triangleq P(A_1 | T_0) ~,&~ \beta_A \triangleq P(A_0 | T_1)
    \end{align}
    where $\alpha_D,\beta_D$ denote the defense system error probabilities and $\alpha_A,\beta_A$ denote the attacker's judge error probabilities.
    Although both the attacker and the defense system aim to reduce these quantities, there is often an inherent trade-off between the type-I and type-II error probabilities. In particular, the defense system may operate under a Neyman-Pearson style criterion, where the false-positive rate $\alpha_D$ is constrained below a tolerable threshold while maximizing detection performance, since rejecting non-harmful responses is a costly error. We also assume that, conditioned on a response's true nature, the attacker-judge and defense decisions are independent. In other words, the attacker cannot causally influence the defense system's decision about an already generated response:
    \begin{align}\label{indepeq}
        (A \perp\!\!\!\perp D) \mid T \;\Leftrightarrow\;  P(A,D | T) = P(A|T)~P(D|T)
    \end{align}

    We now evaluate the probability of event $A_1$, meaning that a received response is selected by the attacker's judge model in Algorithm \ref{algo:attack} as a successful candidate. We start by partitioning the event space using the complementary defense events $D_0$ and $D_1$:
    \begin{align}\label{a1eq}
        P(A_1) = P(A_1 ,D_0) + P(A_1,D_1)
    \end{align}
    We assume that the attacker's judge correctly rejects the predictable refusal or deflection responses returned by detect-and-block, i.e.,
    \begin{align}\label{rejeq}
        P(A_1, D_1) = 0.
    \end{align}
    We can further expand the expression in \eqref{a1eq} with respect to conditional probabilities about the true nature of the response ($T$) as follows:
    \begin{align} \label{a1prob}
        P(A_1)  = \sum_{i=0}^1 P(A_1, D_0 \mid T_i)\,P(T_i)
    \end{align}
    Using the independence condition in \eqref{indepeq} and the notation in \eqref{notationeq} yields:
    \begin{align}
        P(A_1) = q\beta_D(1-\beta_A) + (1-q)(1-\alpha_D)\alpha_A
    \end{align}

    According to the workflow in Algorithm \ref{algo:attack}, an attacker's ultimate success depends on discovering true-positive attack prompts with verifiable effects. Given verification budget $K$, the attacker can optionally verify positive candidates selected in each cycle to avoid false-positive error propagation, or postpone verification until the last cycle. Irrespective of the verification schedule, we consider a cycle successful if it results in at least one verified true-positive attack prompt candidate. Hence, we use the following expression for the ASR per generation cycle. This can also serve as an upper-bound success criterion for frameworks such as \cite{autodan2023, attacktree}, where more than one true-positive candidate may be needed to properly update templates for future attack cycles.
    \begin{align}\label{eq:defasr}
        \mathrm{ASR} &\triangleq P(\mathcal{D}_g^K\ne \emptyset) = 1 - P(\mathcal{D}_g^K = \emptyset)
    \end{align}
    We can then expand the probability term in \eqref{eq:defasr} with respect to the possible sizes of the candidate set, $|\mathcal{C}_g^N|$, using the total probability rule:
    \begin{align}
        \mathrm{ASR} = 1-\sum_{n=0}^N P(\mathcal{D}_g^K = \emptyset~,~|\mathcal{C}^N_g| = n)
    \end{align}
    Since having an empty candidate set implies having no verified candidates, we can further adjust the range in the sum above to
    \begin{align}\label{eq:defasr3}
        \mathrm{ASR} = P(\mathcal{C}^N_g \ne \emptyset) - \sum_{n=1}^N P(\mathcal{D}_g^K = \emptyset~,~|\mathcal{C}^N_g| = n)
    \end{align}

    The attack workflow in Algorithm \ref{algo:attack} permits progressive prompt generation that incorporates decision history about previous prompts. This may correlate judge decisions within the same cycle, so the attempts need not be i.i.d. However, once a candidate set is formed, the verification step evaluates selected prompt-response pairs independently and according to their true nature. We model the per-candidate verification success probability by $\eta_A$, the \textit{positive-predictive value (PPV)} of the attacker's judge, which is computed from the judge error probabilities using Bayes' theorem:
    \begin{align}\label{eq:ppv}
        \eta_A&\triangleq P(T_1 | A_1)\\[8pt]\nonumber
        &=\frac{P(A_1~|~T_1)~P(T_1)}{P(A_1)}\\[8pt]\nonumber
        &=\frac{P(A_1, D_0~|~T_1)~P(T_1)}{P(A_1)}\\[8pt]\nonumber
        &=\frac{q\beta_D(1-\beta_A)}{q\beta_D(1-\beta_A) + (1-q)(1-\alpha_D)\alpha_A}\\[8pt]\nonumber
        &=\left(1 + \frac{1-q}{q}\frac{1-\alpha_D}{\beta_D}\frac{\alpha_A}{1-\beta_A}\right)^{-1}
    \end{align}
    Given the maximum verification budget $K$, we obtain the following expression for each summand in \eqref{eq:defasr3}:
    \begin{align}
        P(\mathcal{D}_g^K = \emptyset~,~|\mathcal{C}^N_g| = n) = P(|\mathcal{C}^N_g| = n)(1-\eta_A)^{\min(n,K)}.
    \end{align}
    We then use the following inequalities
    \begin{align}
        (1-\eta_A)^K \le (1-\eta_A)^{\min(n,K)} \le (1-\eta_A)
    \end{align}
    to substitute the sum terms in \eqref{eq:defasr3} for $n \in \left[1,N\right]$ and derive the following closed-form bounds on ASR:
    \begin{align}\label{eq:asrbounds}
        \mathrm{ASR} &\le P(\mathcal{C}^N_g \ne \emptyset)(1 - (1-\eta_A)^K)\nonumber\\
        \mathrm{ASR} &\ge P(\mathcal{C}^N_g \ne \emptyset)(1 - (1-\eta_A))
    \end{align}

    The remaining term is $P(\mathcal{C}^N_g \ne \emptyset)$, the probability that at least one judge-positive candidate is selected in the cycle.
    The i.i.d. prompt generation case is sufficient for this term to approach one as $N$ grows, but it is not necessary.
    It can approach one as long as the search process remains non-degenerate, meaning that the conditional probability of discovering a judge-positive candidate does not vanish as the attacker continues searching.
    This type of condition is standard in random-search convergence analyses \cite{solis1981minimization} and is discussed in Appendix \ref{app:nondegenerate-search}.
    This condition applies particularly to intermediate attack cycles with more relaxed selection criteria (lower $\tau_g$) and to frameworks that continue searching until such a candidate is found \cite{gcg2023,autodan2023}.
    Under this less restrictive condition,
    \begin{align}\label{eq:largeN}
        \lim_{N\to\infty}P(\mathcal{C}^N_g \ne \emptyset)=1.
    \end{align}
    The familiar expression $1-(1-P(A_1))^N$ is recovered as the homogeneous i.i.d. special case.

    To derive the worst-case bound on ASR from the defense perspective, we evaluate the high-potency prompt regime $q=1$, corresponding to prompts that would trigger harmful responses from the underlying model whenever they are not blocked by the defense.
    This choice is justified by the positive-predictive value term in \eqref{eq:ppv}: for fixed judge error rates, increasing the prior probability of truly harmful responses increases the probability that a judge-positive candidate is a true positive.
    Therefore, the verification component of the ASR bound is maximized at $q=1$, where $\eta_A=1$.
    This is a reasonable choice for worst-case attack prompts that can trigger harmful responses from the underlying model in the absence of any defense system.
    This regime also covers transferred prompts used as high-potency seeds for continued search. A directly replayed prompt is the $N=1$ special case, with $q=1$ as the defense-side worst case. Its success probability is $\beta_D(1-\beta_A)$, or $\beta_D$ under direct verification.
    In this worst-case scenario, the automated judge in the cycle only needs to reliably detect refusals from the target model, which is achievable given the predictable refusal behavior of many current models.
    Combining $\eta_A=1$ with \eqref{eq:largeN} yields
    \begin{align}\label{eq:baseasr}
        \max_q(\mathrm{ASR}) = P(\mathcal{C}^N_g \ne \emptyset)
    \end{align}
    Under the homogeneous i.i.d. approximation, this large-$N$ behavior can be written in the familiar closed form
    \begin{align}\label{eq:baseasr_iid}
        \max_q(\mathrm{ASR}) \approx 1 - (1-\beta_D(1-\beta_A))^N.
    \end{align}
    Thus, given $\beta_D(1-\beta_A)>0$, the attacker can increase its worst-case ASR arbitrarily close to $1$ by increasing the number of model-guided attempts per cycle $N$.
    \begin{remark}[\textbf{Incompleteness of Defense Strategies}]
        Consistent with studies such as \cite{su2024mission_impossible}, which show the theoretical existence of prompts bypassing model alignment, the above analysis implies that detect-and-block defense strategies are incomplete against model-guided attacks capable of automated search in prompt space. Under a non-degenerate search process, their success probability can approach $1$ for a sufficiently large number of attempts $N$, provided that the effective pass-through probability $\beta_D(1-\beta_A)>0$.
        \begin{align}\label{eq:rem1}
            \lim_{N\to\infty} \max_q(\mathrm{ASR}) = 1 .
        \end{align}
    \end{remark}

    \section{Analysis of Defensive Misdirection Strategy} \label{sec:propanalysis}
\begin{figure}[htb!]
        \centering
        \includegraphics[width=\linewidth]{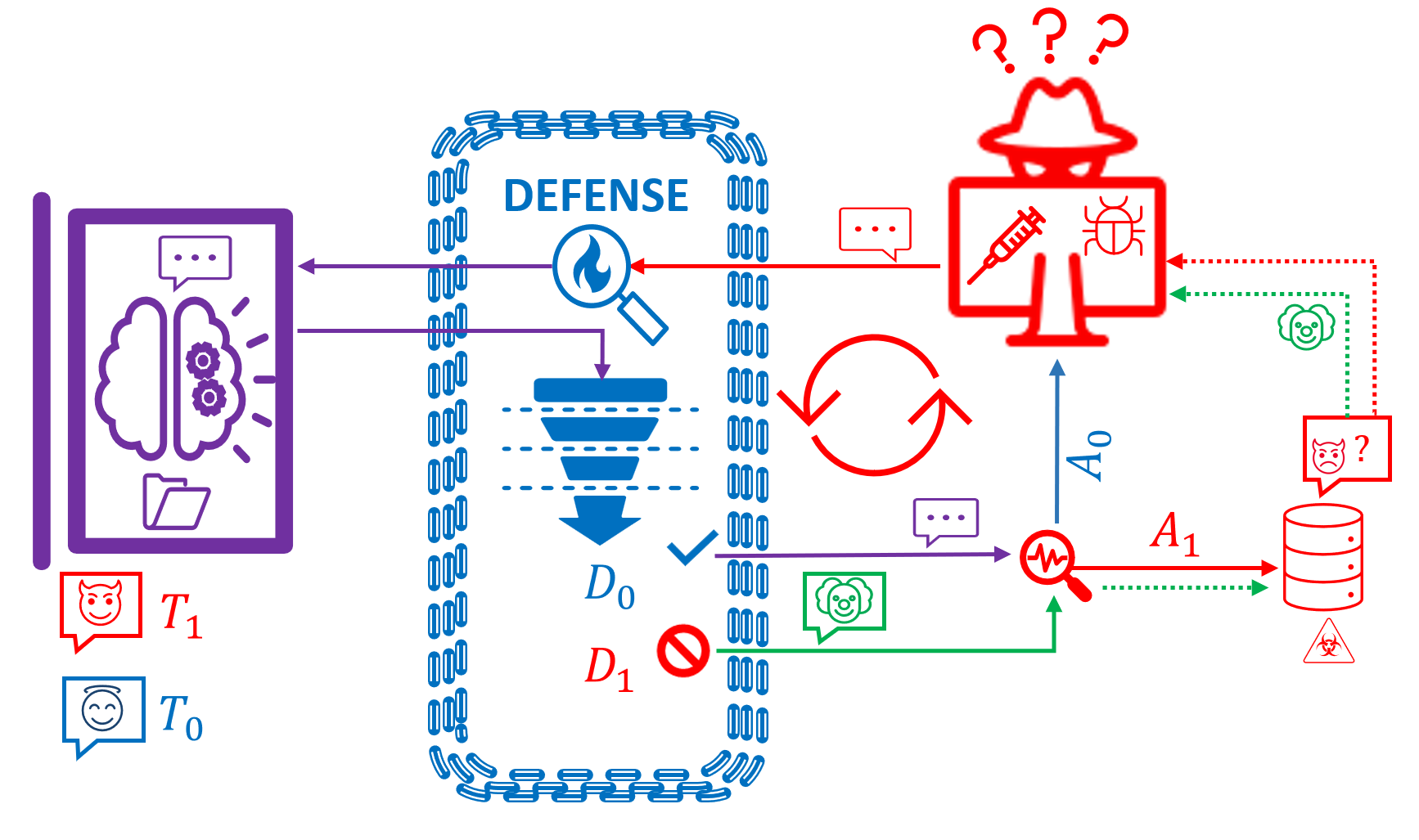}
        \caption{Proposed detect-and-misdirect defense strategy against model-guided attacks}
        \label{fig:proposed}
    \end{figure}

    Our proposed defense strategy, depicted in Figure \ref{fig:proposed}, can be described as \textit{detect-and-misdirect}, complementing current defense strategies against the rising threat of model-guided automated attacks in agentic ecosystems.
    Once it detects a pattern of iterative malicious prompts, the defense mechanism can employ misdirection techniques to produce engaging in-context responses instead of clear and predictable refusal messages.
    The misdirection responses aim to impose a new category of false-positive errors on the attacker's automated judge model, i.e. \textit{misdirection-induced false positives (MI-FP)}, whose probability $\gamma_A$ is defined using the notations introduced in Section \ref{sec:theory}:
    \begin{align}
        \gamma_A \triangleq P(A_1|D_1)
    \end{align}
    This implies that the condition in \eqref{rejeq} no longer holds and
    \begin{align}\label{eq:pa1}
        P(A_1) = \gamma_A P(D_1) + \sum_{i=0}^1 P(A_1, D_0 \mid T_i)\,P(T_i)
    \end{align}
    increases compared to \eqref{a1prob}, where
    \begin{align}
        P(D_1) = q(1-\beta_D)+(1-q)\alpha_D .
    \end{align}
    Misdirection responses are generated only for prompts deemed malicious by the defense system ($D_1$) and are non-functional from the attacker's perspective ($T_0$). This reduces the PPV of candidates discovered by Algorithm~\ref{algo:attack} by increasing the denominator in \eqref{eq:ppv} while leaving its numerator unchanged.
    \begin{align}\label{eq:ppv2}
        &\eta_A = \\\nonumber
        &\left(1 + \frac{(1-q)(1-\alpha_D)\alpha_A}{q\beta_D(1-\beta_A)} + \gamma_A\frac{q(1-\beta_D)+(1-q)\alpha_D}{q\beta_D(1-\beta_A)}\right)^{-1}
    \end{align}
    A significant reduction in $\eta_A$ defeats the purpose of model-guided iterative search, thus forcing the attacker to resort to exhaustive and less practical approaches to verify each prompt.
    Following the same defense-side worst-case regime $q=1$, we evaluate the quantities in \eqref{eq:ppv2} and \eqref{eq:pa1} as follows:

    \begin{align}
        &P(A_1) = \beta_D(1-\beta_A) + \gamma_A(1-\beta_D) \label{eq:pa1q1}\\[8pt]
        &\eta_A =  \left(1+\frac{\gamma_A(1-\beta_D)}{\beta_D(1-\beta_A)}\right)^{-1} \label{eq:ppvq1} ~\mathrm{given}~~ q = 1
    \end{align}

    These expressions can be inserted into the ASR upper bound in \eqref{eq:asrbounds} and then extended for large $N$ according to \eqref{eq:largeN} to obtain the general bound in \eqref{uppereq}.
    Nevertheless, we can derive the following tractable approximation for ASR under the homogeneous i.i.d. prompt generation condition with verification budget $K=1$, which we use for numerical evaluation:
    \begin{align}\label{eq:asrmax}
        &\max_q(\mathrm{ASR})\big|_{K=1} = \\\nonumber
        &\beta_D(1-\beta_A)\sum_{n=0}^{N-1}(\beta_A\beta_D+(1-\gamma_A)(1-\beta_D))^n.
    \end{align}
    This can be used to derive the following simplified worst-case ASR bounds:
    \begin{align}\label{eq:asrbound}
        &\max_q(\mathrm{ASR})\big|_{K=1} \ge \beta_D(1-\beta_A)\nonumber\\
        &\max_q(\mathrm{ASR})\big|_{K=1} < \left(1+\frac{\gamma_A(1-\beta_D)}{\beta_D(1-\beta_A)}\right)^{-1}
    \end{align}
    confirming that the proposed strategy prevents the ASR convergence to $1$ shown in \eqref{eq:rem1} for large numbers of automated attempts $N$, provided that $\gamma_A > 0$.

    \begin{remark}[\textbf{Misdirection Strategy Impact}]
        \label{rem3}
        The following reduced upper bound on the attacker success rate expresses the core strength of the detect-and-misdirect strategy against emerging model-guided automated attack frameworks.
        Integrated with existing defense strategies, replacing predictable refusal messages with in-context misdirection can disrupt the feedback mechanism in such frameworks, forcing attackers to resort to less-practical search approaches with exhaustive and manual verification.
        \begin{align} \label{uppereq}
            \boxed{\lim_{N\to\infty} \max_{q}(\mathrm{ASR}) \le  1-\left(1 + \frac{\beta_D(1-\beta_A)}{\gamma_A(1-\beta_D)}\right)^{-K}}
        \end{align}
        This bound is the consequence of a non-zero misdirection-induced false-positive rate $\gamma_A$, which prevents the asymptotic attack success probability from reaching unity.
        Because blocking and misdirection are triggered by the same detection decisions, they act on the same benign false-positive events.
    \end{remark}

    In a broader context, the proposed detect-and-misdirect strategy can be applied to mitigate other automated attacks, such as black-box fuzzing, in scenarios where the attacker's evaluation feedback is subject to false-positive errors.
    For example, this technique may not be applicable in classic software security scenarios where attackers can directly trigger and verify a desired target outcome (e.g., obtaining a key) without relying on semantic evaluators.
    In contrast, malicious behavior in an agentic AI system can take varied forms whose success may require AI/ML evaluators, which are prone to false-positive errors.
    This distinction is central to model-guided attacks where detecting predictable refusal or deflection behavior is often easier than verifying whether an apparently cooperative response contains a functional malicious output.
    Misdirection complicates the attacker's evaluation task by shifting it from detecting refusal to extracting functional information from responses.
    It can leverage known weaknesses of LLM-based evaluators, which may rely on superficial cues such as tone, structure, and perceived intent rather than strict semantic correctness \cite{chen2024humans, wang2024fair}.

    Attackers may seek to adapt existing model-guided frameworks against misdirection through costly means such as judge ensembling and extra calibration.
    However, these adaptations introduce their own trade-off: stricter judge rules can reduce the misdirection-induced false-positive rate $\gamma_A$, but they may also increase the attacker judge's false-negative rate $\beta_A$, causing the attack loop to discard genuinely harmful candidates.
    Appendix~\ref{app:adaptive-ensemble} evaluates this trade-off through both judge ensembles and a defense-aware PAIR variant.
    Furthermore, attackers may still need more relaxed candidate-selection criteria during early cycles (e.g. lower $\tau_g$ in Algorithm \ref{algo:attack}) to identify weaknesses of the target system with respect to prompt templates that they can make more effective over the next cycles.
    Because relaxed criteria are more prone to accepting false positives, and thus produce larger MI-FP rates $\gamma_A$, detect-and-misdirect can also divert later attack cycles toward ineffective trajectories and resource exhaustion.
    Substantial reduction of $\gamma_A$ in early cycles may require stricter automated evaluation or increased human verification, weakening the automation advantage of model-guided attacks.
    Recent studies further show that LLM-as-a-judge systems exhibit important reliability limitations even for advanced models and may still require expert grounding or human oversight in complex evaluation tasks \cite{szymanski2025limitations, chen2024humans}.
    Consequently, while attackers may temper the effect of misdirection through stronger evaluators or ensembles, reliably filtering misdirection-induced false positives ultimately pushes model-guided attacks toward more costly human verification and exhaustive search.

    Techniques for deceiving adversaries, often referred to as cyber deception, have been studied and applied in other fields of cybersecurity.
    A related decision-theoretic principle appears in secure distributed detection, where probabilistic remapping of sensor reports degrades an eavesdropper's inference while largely preserving authorized detection utility \cite{soosahabi2014optimal}.
    However, cyber deception often involves the deployment of decoy artifacts such as honeytokens and honeypots to divert attacks or analyze them for offensive security research.
    Similarly, the proposed misdirection strategy can make apparent successes less trustworthy to the attacker, since they must spend additional resources distinguishing true bypasses from controlled false engagements.
    This uncertainty may serve as both an operational and psychological deterrent against automated prompt injection and jailbreak attacks, consistent with prior observations that belief in deception mechanisms can deter adversarial behavior \cite{cydec-naval}.

    \section{Contextual Misdirection via Progressive Engagement Against Jailbreak Attacks}\label{sec:cmpe}
    To provide a proof of concept for the proposed detect-and-misdirect strategy, we present a conversational misdirection mechanism termed \textit{Contextual Misdirection via Progressive Engagement (CMPE)}, which we evaluate against automated jailbreak attacks.
    The objective of the CMPE method is to replace predictable refusal responses with responses that appear cooperative and semantically plausible, while subtly disrupting the attacker’s automated evaluation loop without violating the model's policies.

    As illustrated in Algorithm \ref{algo:cmpe}, the CMPE method consists of three components.
    First, a \textit{positive-intent preamble} establishes cooperative framing, which biases automated evaluators toward interpreting the response as desired by the attacker.
    Second, a \textit{context expansion stage} reshapes the original prompt through token-level transformations (e.g., shuffling and lexical injection) and expands it into a coherent but non-operational narrative.
    This stage preserves semantic proximity to the original query while avoiding execution of the malicious intent.
    Third, a \textit{follow-up question} introduces conversational branching to engage the attacker and obscure the evasive intent of the defense mechanism.

    \begin{algorithm}[!]
        \caption{Example of misdirection strategy against jailbreak attacks: Contextual Misdirection via Progressive Engagement (CMPE)}
        \label{algo:cmpe}
        \small
        \begin{algorithmic}[1]
            \Require prompt $p$; lightweight model $M$; expansion budget $N_e$
            \Ensure misdirection response $r = s_{\text{pre}} \,\|\, s_{\text{ctx}} \,\|\, q$

            \State \textbf{(1) Positive-intent preamble}
            \State $s_{\text{pre}} \gets M(\text{make\_positive\_intent\_preamble},\, p)$
            \Comment{1--2 sentences}

            \State \textbf{(2) Safe context via prompt-reshaping}
            \State $w \gets \textsc{Tokens}(\textsc{StripPunct}(p))$
            \State $w \gets \textsc{InjectGlue}(w)$
            \State $w \gets \textsc{Shuffle}(w)$
            \State $\tilde{p} \gets \textsc{Join}(w)$
            \State $s_{\text{ctx}} \gets M(\text{expand\_text},\, \tilde{p},\, N_e)$
            \Comment{expand to $\approx N_e$ words}

            \State \textbf{(3) Follow-up question}
            \State $q \gets M(\text{make\_detailed\_followup\_question},\, p)$
            \Comment{1--2 sentences}

            \State $r \gets s_{\text{pre}} \,\|\, s_{\text{ctx}} \,\|\, q$
            \State \Return $r$
        \end{algorithmic}
        \normalsize
    \end{algorithm}

    Looking at the example in Figure \ref{fig:cmpe}, CMPE shares similarities with techniques studied in computational persuasion, where responses are designed to influence beliefs in interacting parties \cite{computational_persuasion}.
    However, when used as a defense mechanism, CMPE is designed to disrupt adversarial evaluation mechanisms in automated jailbreak attacks by introducing controlled misdirection.
    In particular, the generated responses exploit known limitations of LLM-based judges, which can rely on heuristic cues such as tone, structure, and perceived intent rather than strict semantic correctness \cite{zheng2023judging}.

    The results in Section \ref{sec:num} demonstrate how CMPE can significantly increase misdirection-induced false positives in leading evaluation models and disrupt the automated jailbreak frameworks presented in \cite{pair2023, llmfuzzer2024}.
    Appendix \ref{app:ycmpe} shows that CMPE is safer and more effective than instructing an abliterated (refusal-suppressed) model to generate misdirection responses.
    The CMPE method can also be used to create training datasets for integrating the detect-and-misdirect strategy as a learned technique in agentic security components.

    \section{Simulation \& Numerical Results} \label{sec:num}
    In this section, we estimate judge error rates for malicious and CMPE responses, then use the per-sample estimates to compute maximum ASR bounds for detect-and-block and detect-and-misdirect across different attacker--defender judge configurations.
    \subsection{Jailbreak Prompts Dataset}\label{sec:dset}
    We evaluate our framework using a subset of 500 high-risk jailbreak prompts drawn from the \textit{AdvBench dataset} introduced in \cite{gcg2023}.
    These prompts represent a diverse set of illicit and harmful intent queries commonly used as seed prompts for red-teaming aligned language models.
    The dataset is publicly available\footnote{\url{https://github.com/llm-attacks/llm-attacks}} and has been adopted in several jailbreak benchmarks and attack evaluations, including PAIR, AutoDAN, and JailbreakBench \cite{pair2023, autodan2023, jailbreakbench2024}.
    \subsection{Response Generation Model}\label{sec:resp-gen}
    We use the NeuralDaredevil-8B-abliterated model\footnote{\url{https://huggingface.co/mlabonne/NeuralDaredevil-8B-abliterated}}, a lightweight refusal-suppressed model based on the Meta LLaMA-3 architecture, as the response generation engine for two purposes:
    \subsubsection{Malicious response generation}
    For each AdvBench prompt, we generate an explicitly malicious response to estimate judge false-negative rates. Since these direct responses are easy to verify manually, they provide a conservative estimate of judge false-negative rates relative to obfuscated or role-play attack responses. For quality control, each response is evaluated five times by each of the six judges in Section \ref{sec:judgemodels}. We retained 453 samples with an average normalized score above 0.75 on at least four judges and manually modified and retested the remaining 47 to meet this criterion.
    \subsubsection{Misdirection generation}\label{subsec:misdgen}
    Since aligned models often refuse to process prompts with harmful context, we use the same abliterated model in CMPE response generation for context reshaping and follow-up engagement according to Algorithm~\ref{algo:cmpe}, with a context expansion budget of $N_e=200$ words.
    \subsection{Judge Models}\label{sec:judgemodels}
    We evaluate six judge configurations spanning rubric-based LLM judges and trained classifiers. StrongREJECT (SR) \cite{sr2024} and PAIR \cite{pair2023} score response compliance and harmfulness in $[0,5]$ and $[1,10]$ ranges, respectively. HB-FT-LLaMA2-13B (HarmBench Fine-Tuned) \cite{harmbench}, GPTFuzz-RoBERTa \cite{yu2023gptfuzzer}, and Llama-Guard-3-8B \cite{llamaguard3modelcard} perform binary classification of harmful content or unsafe prompt-response behavior. For GPTFuzz-RoBERTa, confidence scores are incorporated into normalized harmfulness scores. We omit keyword-based refusal detectors because CMPE avoids predictable refusal wording.
    All local models were deployed on a pair of interconnected {NVIDIA} DGX Spark\texttrademark~systems.\footnote{Each {NVIDIA} DGX Spark\texttrademark~system is equipped with a GB10 Grace Blackwell\texttrademark~superchip and 128\,GB of coherent memory.}
    We used GPT-OSS-120B and LLaMA-4-Scout-17B reasoning models to implement rubric-based LLM judges.
    \subsection{Estimating Judge Error Rates} \label{sec:err}
    For each of the 500 prompts, we evaluate the paired malicious and CMPE {responses} 10 times per judge, yielding 5{,}000 outputs per response type and judge. Output scores are normalized to $[0,1]$ to be comparable with binary classifier outputs, with larger values indicating {stronger judgments of harmfulness}. The full score distributions are reported in Figures \ref{fig:vplot-malscores} and \ref{fig:vplot-scores} in Appendix \ref{sim_app}. They indicate that CMPE responses {receive} high scores for a large fraction of prompts. We average the 10 normalized outputs for each {prompt-response pair} to obtain an attacker judge-positive probability proxy. {The complement of this value} for malicious responses estimates the per-sample false-negative (FN) rate, while {the value itself} for CMPE responses estimates the misdirection-induced false-positive (MI-FP) rate. Table \ref{tabbg} reports the per-judge averages of these quantities across all 500 samples.
    \subsection{ASR Evaluation \& Comparison}
    \label{sec:asr_eval}
    Using the per-sample error estimates from Section~\ref{sec:err}, we compute and average the upper bound on maximum ASR for each attacker--defender judge configuration under the homogeneous i.i.d. special case in \eqref{eq:asrmax}. Each bound uses the corresponding per-sample estimates of $\beta_D$, $\beta_A$, and $\gamma_A$, with $\gamma_A=0$ for detect-and-block. The final ASR bound values are calculated by averaging 500 estimated per-sample ASR bounds instead of using the aggregated rates in Table \ref{tabbg}. For $N=100$ and $K=1$, detect-and-block often yields ASR bounds near $1.0$, whereas all CMPE bounds are at most $0.165$ and {27 of the 30 judge configurations are} below $0.1$. The reduction reaches approximately two orders of magnitude, for example{,} from $0.413$ to $0.004$. Full matrices are reported in Tables \ref{tab:asr_base} and \ref{tab:asr_def} in Appendix \ref{sim_app}. The increase in $\gamma_A$ lowers the attacker's positive predictive value (PPV) and degrades candidate selection, demonstrating the mechanism predicted in Section \ref{sec:propanalysis}.

    \section{Empirical Evaluation Against Automated Attack Frameworks}
    \label{sec:e2e_eval}
    {We conduct end-to-end adversarial evaluations using three representative model-guided attack frameworks: {GPTFuzz}~\cite{yu2023gptfuzzer,llmfuzzer2024}, {PAIR}~\cite{pair2023}, and AutoDAN-Turbo~\cite{autodanturbo2025}.}
    These experiments emulate an agentic security setting in which a vulnerable victim model is protected by a content-moderation defense model that either blocks detected harmful responses or replaces them with CMPE misdirection.
    {The goal is to measure verified ASR when the full attack framework interacts with the defended victim system under these two defense strategies.}
    \begin{table*}[h]
        \centering
        \setlength{\tabcolsep}{4pt}
        \renewcommand{\arraystretch}{1.2}
        \caption{Estimated average judge error rates across all $500$ samples. These aggregate rates summarize judge behavior.}
        \begin{tabular}{lcccccc}
            \toprule
            \textbf{Estimated Ave. Error}
            & \textbf{\shortstack{SR-OSS-120B}}
            & \textbf{\shortstack{SR-Scout-30B}}
            & \textbf{\shortstack{PAIR-OSS-120B}}
            & \textbf{\shortstack{HB-FT-LLaMA2-13B}}
            & \textbf{\shortstack{LLaMA-Guard-3-8B}}
            & \textbf{\shortstack{GPTFuzz-RoBERTa}} \\
            \midrule
            $\hat{\beta}$ (FN)
            & \basecell{0.010911}
            & \basecell{0.084889}
            & \basecell{0.023289}
            & \basecell{0.054089}
            & \basecell{0.068889}
            & \basecell{0.205822} \\

            $\hat{\gamma}_A$ (MI-FP)
            & \defcell{0.6272}
            & \defcell{0.7380}
            & \defcell{0.8515}
            & \defcell{0.4276}
            & \defcell{0.9804}
            & \defcell{0.7145} \\
            \bottomrule
        \end{tabular}
        \label{tabbg}
    \end{table*}
    \begin{table*}[!t]
        \centering
        \small
        \setlength{\tabcolsep}{4pt}
        \renewcommand{\arraystretch}{1.2}
        \caption{PAIR end-to-end results over 50 prompts with a maximum of 50 iterations per prompt. Positive and negative exits denote framework-reported termination outcomes. MI-FP denotes exits caused by CMPE responses accepted as successful by the attacker judge. ASR is computed as validated true positives (TP) divided by the total number of prompts.}
        \begin{tabular*}{\textwidth}{@{\extracolsep{\fill}}llcccccccc@{}}
            \toprule
            \textbf{Victim} & \textbf{Defense} & \textbf{Total}
            & \textbf{\shortstack{Positive Exits}}
            & \textbf{\shortstack{Negative Exits}}
            & \textbf{\shortstack{MI-FP}}
            & \textbf{FP}
            & \textbf{TP}
            & \textbf{ASR}
            & \textbf{\shortstack{Avg. Iter.}} \\
            \midrule
            Vicuna      & LLaMA-Guard-3-8B        & 50 &  8 & 42 &  0 & 3 & 5 & \basecell{0.100} & 45.4 \\
            Vicuna      & LLaMA-Guard-3-8B + CMPE & 50 & 43 &  7 & 43 & 0 & 0 & \basecell{0.000} & 20.4 \\
            \midrule
            Abliterated & LLaMA-Guard-3-8B        & 50 & 12 & 38 &  0 & 7 & 5 & \basecell{0.100} & 41.4 \\
            Abliterated & LLaMA-Guard-3-8B + CMPE & 50 & 47 &  3 & 47 & 0 & 0 & \basecell{0.000} & 15.8 \\
            \bottomrule
        \end{tabular*}
        \label{tab:pair_e2e}
    \end{table*}
    \begin{table*}[!t]
        \centering
        \small
        \setlength{\tabcolsep}{4pt}
        \renewcommand{\arraystretch}{1.2}
        \caption{GPTFuzz end-to-end results over 50 prompts with a maximum of 50 iterations per prompt. Positive and negative exits denote framework-reported termination outcomes. MI-FP denotes exits caused by CMPE responses accepted as successful by the attacker judge. ASR is computed as validated true positives (TP) divided by the total number of prompts.}
        \begin{tabular*}{\textwidth}{@{\extracolsep{\fill}}llcccccccc@{}}
            \toprule
            \textbf{Victim} & \textbf{Defense} & \textbf{Total}
            & \textbf{\shortstack{Positive Exits}}
            & \textbf{\shortstack{Negative Exits}}
            & \textbf{\shortstack{MI-FP}}
            & \textbf{FP}
            & \textbf{TP}
            & \textbf{ASR}
            & \textbf{\shortstack{Avg. Iter.}} \\
            \midrule
            Vicuna      & LLaMA-Guard-3-8B        & 50 & 23 & 27 &  0 & 13 & 10 & \basecell{0.200} & 37.5 \\
            Vicuna      & LLaMA-Guard-3-8B + CMPE & 50 & 48 &  2 & 44 &  4 &  0 & \basecell{0.000} &  6.8 \\
            \midrule
            Abliterated & LLaMA-Guard-3-8B        & 50 & 22 & 28 &  0 & 12 & 10 & \basecell{0.200} & 34.0 \\
            Abliterated & LLaMA-Guard-3-8B + CMPE & 50 & 48 &  2 & 42 &  5 &  1 & \basecell{0.020} &  6.4 \\
            \bottomrule
        \end{tabular*}
        \label{tab:gptfuzz_e2e}
    \end{table*}
    \begin{table*}[!t]
        \centering
        \small
        \setlength{\tabcolsep}{4pt}
        \renewcommand{\arraystretch}{1.2}
        \caption{AutoDAN-Turbo end-to-end results over 50 prompts with a maximum of 20 epochs per prompt. Positive and negative exits denote framework-reported termination outcomes. MI-FP denotes exits caused by CMPE responses accepted as successful by the attacker judge. ASR is computed as validated true positives (TP) divided by the total number of prompts.}
        \begin{tabular*}{\textwidth}{@{\extracolsep{\fill}}llcccccccc@{}}
            \toprule
            \textbf{Victim} & \textbf{Defense} & \textbf{Total}
            & \textbf{\shortstack{Positive Exits}}
            & \textbf{\shortstack{Negative Exits}}
            & \textbf{\shortstack{MI-FP}}
            & \textbf{FP}
            & \textbf{TP}
            & \textbf{ASR}

            & \textbf{\shortstack{Avg. Epochs}} \\
            \midrule
            Vicuna      & LLaMA-Guard-3-8B        & 50 &  6 & 44 &  0 & 3 & 3 & \basecell{0.060} & 18.8 \\
            Vicuna      & LLaMA-Guard-3-8B + CMPE & 50 &  5 & 45 &  5 & 0 & 0 & \basecell{0.000} & 19.2 \\
            \midrule
            Abliterated & LLaMA-Guard-3-8B        & 50 & 13 & 37 &  0 & 5 & 8 & \basecell{0.160} & 16.3 \\
            Abliterated & LLaMA-Guard-3-8B + CMPE & 50 & 26 & 24 & 26 & 0 & 0 & \basecell{0.000} & 13.0 \\
            \bottomrule
        \end{tabular*}
        \label{tab:turbo_e2e}
    \end{table*}

    \subsection{Experimental Configuration}
    {Each attack framework was evaluated on the first 50 adversarial goals in AdvBench~\cite{gcg2023}, using the same goal set across frameworks. GPTFuzz and PAIR used a maximum budget of 50 iterations per prompt, while AutoDAN-Turbo used a maximum of 20 epochs.}
    Two victim models were used: {Vicuna-13b-v1.5}, an aligned target used in prior evaluations of these frameworks, and NeuralDaredevil-8B-abliterated, the same refusal-suppressed Llama 3 model discussed in Section \ref{subsec:misdgen}.

    {GPT-3.5-Turbo and Gemma-1-7B, respectively, served as the attacker generator model for GPTFuzz and AutoDAN-Turbo, following their public artifacts. We also used NeuralDaredevil-8B-abliterated as the attack generator model for PAIR to avoid attack disruptions due to underlying model alignment.}
    GPTFuzz-RoBERTa~\cite{yu2023gptfuzzer} served as the success judge for
    GPTFuzz, while GPT-OSS-120B used the PAIR judge instruction from
    \cite{pair2023}. For AutoDAN-Turbo, GPT-OSS-120B served as the scorer
    backend while preserving the original 1--10 rubric, four score ranges,
    and stopping rule.
    Local defense-side and attack-side components were hosted on separate NVIDIA DGX Spark systems.

    {On the defense side, Llama-Guard-3-8B acted as the content-moderation judge, and NeuralDaredevil-8B-abliterated was used for CMPE response generation as described in Section~\ref{subsec:misdgen}.}
    Together, the two victim models allow us to test the defense under both aligned and refusal-suppressed target settings.
    The {detect-and-block} strategy routes each victim response through Llama-Guard-3-8B; responses classified as harmful are replaced with a refusal message.
    The {detect-and-misdirect} strategy instead invokes the CMPE reshaping algorithm (Algorithm~\ref{algo:cmpe}) to generate a misdirected response that appears responsive to the attacker's query while containing no genuinely harmful content.

    \subsection{Final Response Validation}
    {Positive exits caused by CMPE were recorded as MI-FPs after confirming that the controlled response contained no harmful output. All other positive exits were independently classified by Claude Sonnet 4.6 and ChatGPT 5.5 as either false positives or true positives. Disagreements between the two judges were resolved through manual inspection of the final prompt-response pair. A non-CMPE response was counted as a true positive only if it contained genuinely harmful and actionable content; otherwise, it was counted as a false positive.}
    Each run ends either as a \textbf{Positive Exit}, where the attacker's judge claims success, or a \textbf{Negative Exit}, where the framework terminates without a successful claim within its iteration or epoch budget.
    Among positive exits, \textbf{MI-FP} denotes controlled CMPE misdirection responses accepted as successful, \textbf{FP} denotes other safe or invalid responses incorrectly accepted by the attacker judge, and \textbf{TP} denotes responses confirmed as genuinely harmful by final validation.
    {\textbf{Avg.\,Iter.} is the mean number of iterations to termination over all 50 prompts. Table~\ref{tab:turbo_e2e} reports the corresponding mean number of epochs for AutoDAN-Turbo.}

    \subsection{Results and Interpretation}

    {Tables~\ref{tab:gptfuzz_e2e}, \ref{tab:pair_e2e}, and~\ref{tab:turbo_e2e} present the end-to-end results for GPTFuzz, PAIR, and AutoDAN-Turbo, respectively.}
    Under detect-and-block, GPTFuzz produced 23 and 22 positive exits on Vicuna and the abliterated model, respectively.
    Final validation confirmed 10/50~(20\%) true positives in both cases, while 13 and 12 positive exits were false positives.
    The remaining 27--28 runs ended as negative exits, terminating after 34.0--37.5 iterations on average.
    PAIR was more conservative, producing 8 and 12 positive exits, of which 5/50~(10\%) were validated true positives on both victim models, with 3 and 7 false positives, respectively; the remaining 38--42 runs were negative exits and terminated after 41.4--45.4 iterations on average.

    Under detect-and-misdirect, GPTFuzz produced 48 positive exits on both victim models.
    On Vicuna, 44/48 positive exits were MI-FPs and the remaining 4 were ordinary false positives, yielding 0 validated true positives.
    On the abliterated model, 42/48 positive exits were MI-FPs, 5 were ordinary false positives, and only 1/50~(2\%) was a validated true positive.
    Only 2 runs ended as negative exits in both cases, and termination occurred much earlier, at 6.4--6.8 iterations on average.
    For PAIR, detect-and-misdirect produced 43 and 47 positive exits on Vicuna and the abliterated model, respectively.
    All positive exits were MI-FPs, resulting in 0 validated true positives on both targets.
    The remaining 7 and 3 runs ended as negative exits, with average termination at 15.8--20.4 iterations.

    {AutoDAN-Turbo shows a different pattern under its graded scoring rubric. Under detect-and-block, it produced 6 and 13 positive exits, with 3/50~(6\%) and 8/50~(16\%) validated true positives on Vicuna and the abliterated model, respectively. With CMPE, validated ASR fell to 0 on both targets, and all 5 positive exits on Vicuna and all 26 on the abliterated model were MI-FPs. On Vicuna, 45/50 runs ended as negative exits and the average epochs remained near the maximum budget at 19.2. The predominance of negative exits is consistent with CMPE degrading the scorer feedback used for strategy selection, leaving the search with fewer effective candidates. On the abliterated model, 24/50 runs ended as negative exits, while 26 runs terminated on MI-FPs and the average epochs decreased from 16.3 to 13.0. Thus, CMPE prevents validated success through a target-dependent combination of negative exits and MI-FP termination.}

    {For GPTFuzz and PAIR, misdirection converts many attacker-claimed successes into MI-FPs, reducing verified ASR from 20\% to 0--2\% for GPTFuzz and from 10\% to 0\% for PAIR. Their reduced iteration counts reflect premature termination on controlled non-harmful responses rather than improved attack efficiency. AutoDAN-Turbo reaches 0 verified ASR through a target-dependent combination of MI-FP termination and negative exits. Across all three frameworks, attacker-side positive exits no longer reliably correspond to validated harmful outputs.}

    \section{Related Work on Active Defenses}
    \label{sec:related-misdirection}
    Recent studies have empirically explored active defenses that exploit the dependence of automated attackers on LLM interpretation and response evaluation.
    Mantis uses defensive prompt injection against LLM-driven cyberattack agents by planting crafted inputs into system responses, causing the attacker's LLM agent to disrupt its own operation or, in active variants, compromise the attacker's environment \cite{pasquini2024hacking}.
    This line of work is related to our use of adversarial misdirection, but its target is a tool-using cyberattack agent rather than the automated response judge used by model-guided jailbreak frameworks.

    Another related work, \textit{Proactive Defense Against LLM Jailbreak}, introduces ProAct as a proactive defense framework for iterative jailbreak attacks \cite{zhao2025proactive,zhao2025proactive_openreview}.
    ProAct generates benign ``spurious responses'' that appear to satisfy harmful requests, causing iterative jailbreak methods to misinterpret the response as a successful jailbreak and terminate prematurely.
    Our work shares the high-level intuition that apparent success can be safer for the defender than predictable refusal, but differs in its analytical focus and security abstraction.
    Rather than treating misdirection primarily as an empirical proactive-defense mechanism, we formalize the attacker's response evaluation process as the central point of failure, introduce misdirection-induced false positives as an explicit attacker-side error term, and derive ASR bounds in terms of defense false-negatives, attacker judge false-negatives, verification budget, and misdirection-induced false-positive rate.
    This analysis explains why detect-and-misdirect differs from simple refusal or termination: refusal provides a negative search signal, while misdirection corrupts the attacker's candidate selection by lowering the positive predictive value of its judge.
    We quantify judge-ensemble trade-offs using the simulated error rates and evaluate defense-aware adaptation with a repeated-scoring PAIR variant. We also evaluate CMPE through sample-averaged ASR simulations and end-to-end experiments with PAIR, GPTFuzz, and AutoDAN-Turbo, reporting validated TP, FP, and MI-FP outcomes.

    \section{Concluding Discussion}
    \label{sec:concdisc}
    This paper studies the security dynamics of model-guided automated attacks
    against LLM-based agentic systems, with particular focus on the role of automated
    evaluators in scaling jailbreak and prompt-injection attempts.
    We showed that conventional detect-and-block defense strategies can still leave attackers with an asymptotic advantage when repeated automated attempts are available.
    The key reason is that blocking responses provide structured feedback that can be used by model-guided attack frameworks to refine candidate prompts.
    To address this limitation, we introduced a detect-and-misdirect defense strategy that degrades the attacker’s automated evaluation process by introducing controlled, non-operational responses that increase misdirection-induced false positives.
    Our probabilistic analysis shows that this mechanism can bound the attacker success rate even as the query budget grows, including under more general non-degenerate search conditions beyond homogeneous i.i.d. prompt generation.
    We further instantiated the idea through Contextual Misdirection via Progressive Engagement (CMPE) and showed, across multiple automated evaluators and end-to-end frameworks such as PAIR, GPTFuzz, and AutoDAN-Turbo, that misdirection can substantially reduce verified attack success and induce false-positive termination.

    \subsection{Limitations}
    \label{sec:limitations}

    The scope of this work leaves several important considerations for future analysis and deployment.
    The probabilistic derivation adopts a per-cycle analytical abstraction with a constrained verification budget.
    This abstraction is useful for exposing the per-cycle scaling behavior of model-guided attacks, but it does not fully model how misdirection propagates across multiple dependent attack cycles, where false-positive candidates can influence future prompt templates and attacker heuristics.
    In practice, this multi-cycle effect may strengthen the practical impact of misdirection, but it requires a more detailed sequential model of attacker adaptation.

    Our empirical evaluation focuses primarily on jailbreak-style benchmarks and well-known attack frameworks.
    Although the proposed mechanism is motivated by broader automated attacks in agentic AI systems, including prompt injection, further evaluation is needed in full tool-use and multi-agent environments.
    We also did not conduct a dedicated study of the collateral effects of misdirection on benign human users.
    The strategy may be particularly suitable for fully agentic environments where suspicious interactions can be handled without directly confusing a human user.
    In chatbot settings, however, the frequency and conditions under which misdirection should be triggered depend on the confidence and policy judgment of the defending system.
    Clear refusal messages may still be appropriate, and sometimes necessary, as warnings or policy explanations for benign users.

    \subsection{Future Work}
    \label{sec:future-work}

    Future work should study how detect-and-misdirect can be integrated into
    realistic defense architectures.
    In this paper, CMPE is implemented as a separate response-generation mechanism activated after malicious behavior is
    detected.
    A natural next step is to investigate whether misdirection can be incorporated into model-level safety tuning, security reinforcement, or AI
    firewall policies.
    However, directly teaching large general-purpose models when
    to misdirect may be difficult and could increase the risk of unintended
    hallucination or confusing responses.
    A more controlled deployment option is to
    use a separate lightweight model or policy module that generates either refusal
    or misdirection responses depending on the confidence, context, and severity of
    the detected threat.

    An important direction is to study detect-and-misdirect defenses in richer agentic
    settings where the target system interacts with tools, memory, external
    documents, and other agents.
    Such environments create more complex attack
    surfaces, but also provide additional opportunities for controlled response
    shaping.
    Motivated by our analysis of defensive misdirection, recent work applies a
    deception-based defense to black-box extraction of inference-time agent
    harnesses while largely preserving agent utility \cite{cui2026agentharness}.
    This application suggests extending our probabilistic model from binary attack
    success to graded objectives such as extraction fidelity.

    Finally, future work should evaluate the effect of misdirection in human-facing
    security studies, such as controlled CTF-style experiments.
    Such studies could
    measure whether misdirection changes attacker behavior, confidence, and
    resource allocation in non-automated or semi-automated scenarios.
    This would
    also clarify the relationship between detect-and-misdirect and broader cyber
    deception mechanisms, including whether the belief that a target system may deploy misdirection has a deterrent effect comparable to that observed for other cyber-deception mechanisms \cite{cydec-naval}.

    \bibliographystyle{IEEEtran}
    \bibliography{llmaad-929}
    \appendices

    \section{Non-Degenerate Search Condition}
    \label{app:nondegenerate-search}

    The limit in \eqref{eq:largeN} does not require the prompt attempts inside a
    generation cycle to be independent and identically distributed (i.i.d.).
    It only requires that the search process maintains a non-vanishing conditional
    probability of producing a judge-positive candidate as attempts continue.
    This type of non-degeneracy condition is standard in random-search convergence
    analysis \cite{solis1981minimization}.

    Let $A_{1,i}$ denote the event that the $i$-th attempt in generation cycle $g$
    is selected as a candidate by the attacker's judge, i.e., $s_i \ge \tau_g$ in
    Algorithm~\ref{algo:attack}.
    Let $\mathcal{C}_g^i$ denote the candidate set after the first $i$ attempts in
    that cycle, with $\mathcal{C}_g^0=\emptyset$.
    Suppose that before the first candidate is found, there exists some
    $\epsilon>0$ such that
    \begin{align}
        P(A_{1,i}\mid \mathcal{C}_g^{i-1}=\emptyset) \ge \epsilon
    \end{align}
    for all attempts $i$.
    Then the probability of producing no candidate after $N$ attempts satisfies
    \begin{align}
        P(\mathcal{C}_g^N=\emptyset)
        &=
        \prod_{i=1}^{N}
        \left(1-P(A_{1,i}\mid \mathcal{C}_g^{i-1}=\emptyset)\right)\\
        &\le (1-\epsilon)^N .
    \end{align}
    Therefore,
    \begin{align}
        \lim_{N\to\infty}P(\mathcal{C}_g^N=\emptyset)=0,
        \qquad
        \lim_{N\to\infty}P(\mathcal{C}_g^N\ne\emptyset)=1 .
    \end{align}
    This proves that the convergence in \eqref{eq:largeN} holds under a condition less restrictive than assuming i.i.d. prompt attempts.
    The homogeneous i.i.d. expression $1-(1-P(A_1))^N$ is recovered as a special case
    when each attempt has the same candidate-selection probability and attempts are
    independent.

    \section{Extended Simulation Results}\label{sim_app}
\begin{figure}[h!]
        \centering
        \includegraphics[width=\linewidth]{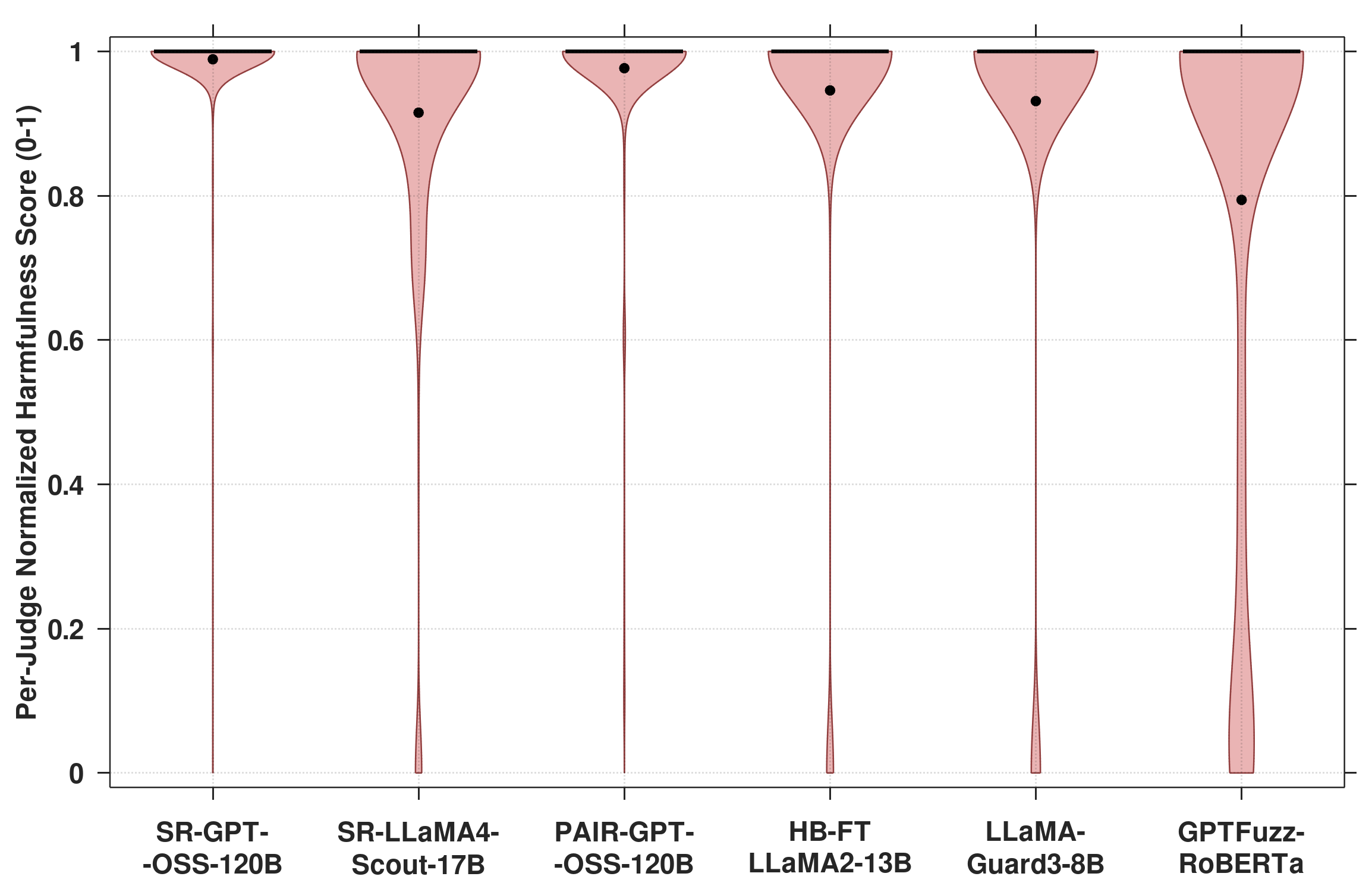}
        \caption{Per-judge normalized harmfulness score distributions for simulated malicious response samples.}
        \label{fig:vplot-malscores}
    \end{figure}

\begin{figure}[h!]
        \centering
        \includegraphics[width=\linewidth]{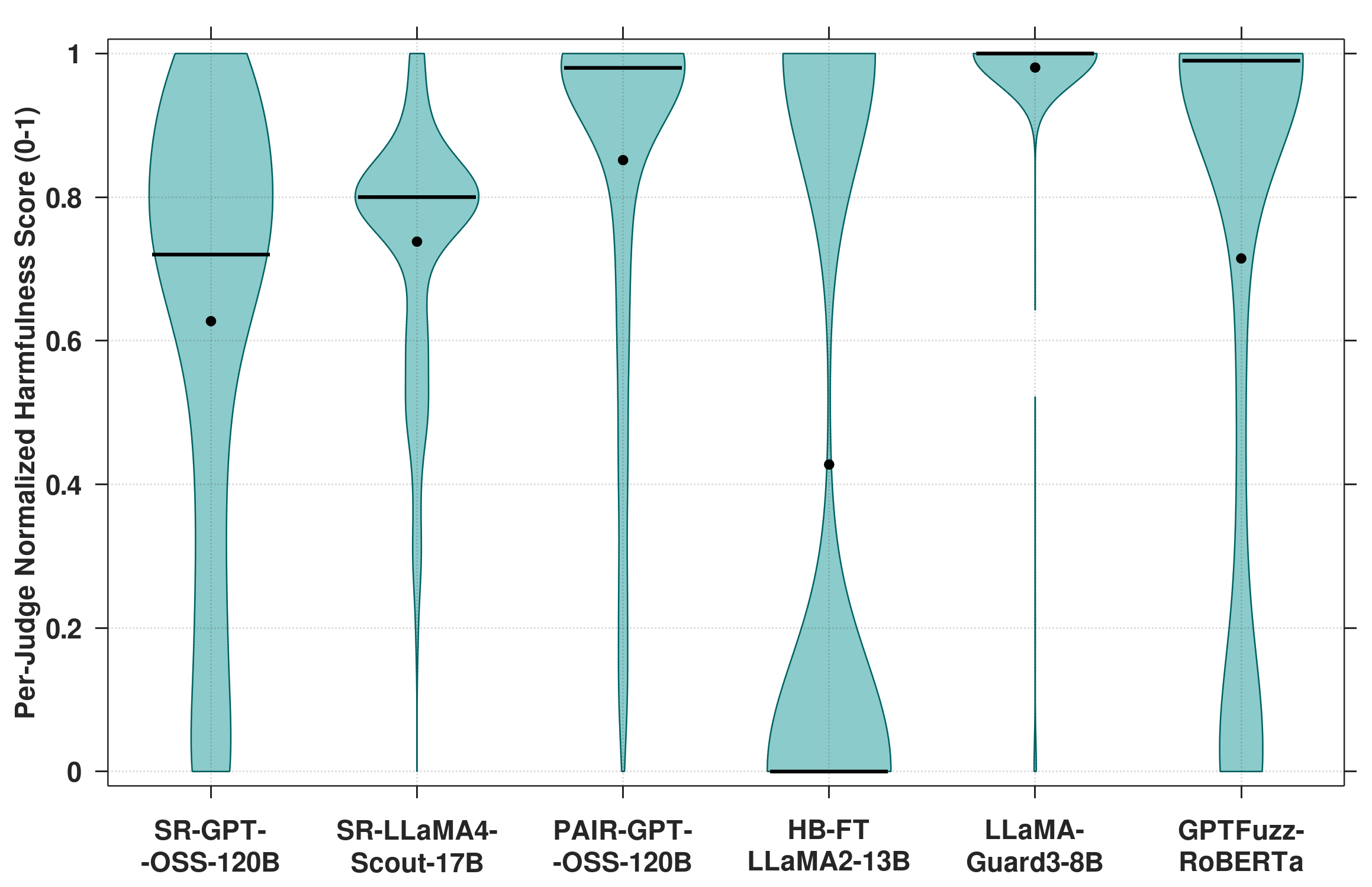}
        \caption{Per-judge normalized harmfulness score distributions for CMPE responses.}
        \label{fig:vplot-scores}
    \end{figure}

    This appendix provides the score distributions and complete defender--attacker ASR matrices supporting Section~\ref{sec:num}. Figures~\ref{fig:vplot-malscores} and~\ref{fig:vplot-scores} show the normalized judge-score distributions for the simulated malicious and CMPE responses, respectively. Tables~\ref{tab:asr_base} and~\ref{tab:asr_def} report the sample-averaged upper bounds on maximum ASR for each judge configuration under detect-and-block and detect-and-misdirect with $N=100$ and $K=1$. Each entry is computed from per-sample error estimates and then averaged over the $500$ prompts rather than from the aggregate rates in Table~\ref{tabbg}.
    \begin{table*}[htb!]
        \centering
        \setlength{\tabcolsep}{4pt}
        \renewcommand{\arraystretch}{1.2}
        \caption{Sample-averaged upper bound on maximum ASR under the detect-and-block strategy for $N=100$ homogeneous i.i.d. attack attempts with $K=1$. For each prompt, ASR is computed using per-sample judge error estimates with $\gamma_A=0$. The table reports the average across all $500$ prompts.}
        \begin{tabular}{lcccccc}
            \toprule
            \textbf{Defender $\downarrow$ / Attacker $\rightarrow$}
            & \textbf{SR-OSS-120B} & \textbf{SR-Scout-30B}
            & \textbf{PAIR-OSS-120B} & \textbf{HB-FT-LLaMA2-13B}
            & \textbf{LLaMA-Guard-3-8B} & \textbf{GPTFuzz-RoBERTa} \\
            \midrule
            \textbf{SR-OSS-120B}
            & \basecell{0.4133} & \basecell{0.5248}
            & \basecell{0.5399} & \basecell{0.4757}
            & \basecell{0.4460} & \basecell{0.5196} \\

            \textbf{SR-Scout-30B}
            & \basecell{0.9998} & \basecell{0.9741}
            & \basecell{0.9998} & \basecell{0.9997}
            & \basecell{0.9994} & \basecell{0.9981} \\

            \textbf{PAIR-OSS-120B}
            & \basecell{0.8689} & \basecell{0.8635}
            & \basecell{0.6037} & \basecell{0.7558}
            & \basecell{0.5620} & \basecell{0.7858} \\

            \textbf{HB-FT-LLaMA2-13B}
            & \basecell{0.9938} & \basecell{0.9930}
            & \basecell{0.9898} & \basecell{0.1488}
            & \basecell{0.9898} & \basecell{0.9507} \\

            \textbf{LLaMA-Guard-3-8B}
            & \basecell{0.9986} & \basecell{0.9968}
            & \basecell{0.9960} & \basecell{0.9979}
            & \basecell{0.001} & \basecell{0.9473} \\

            \bottomrule
        \end{tabular}
        \label{tab:asr_base}
    \end{table*}
    \begin{table*}[htb!]
        \centering
        \setlength{\tabcolsep}{4pt}
        \renewcommand{\arraystretch}{1.2}
        \caption{Sample-averaged upper bound on maximum ASR under the CMPE-based detect-and-misdirect strategy for $N=100$ homogeneous i.i.d. attack attempts with $K=1$. For each prompt, ASR is computed using per-sample judge error estimates including $\gamma_A$, and the table reports the average across all $500$ prompts.}
        \begin{tabular}{lcccccc}
            \toprule
            \textbf{Defender $\downarrow$ / Attacker $\rightarrow$}
            & \textbf{SR-OSS-120B} & \textbf{SR-Scout-30B}
            & \textbf{PAIR-OSS-120B} & \textbf{HB-FT-LLaMA2-13B}
            & \textbf{LLaMA-Guard-3-8B} & \textbf{GPTFuzz-RoBERTa} \\
            \midrule
            \textbf{SR-OSS-120B}
            & \basecell{0.0035} & \basecell{0.0053}
            & \basecell{0.0049} & \basecell{0.0071}
            & \basecell{0.0027} & \basecell{0.0053} \\

            \textbf{SR-Scout-30B}
            & \basecell{0.1237} & \basecell{0.0491}
            & \basecell{0.0942} & \basecell{0.1650}
            & \basecell{0.0740} & \basecell{0.0841} \\

            \textbf{PAIR-OSS-120B}
            & \basecell{0.0275} & \basecell{0.0229}
            & \basecell{0.0066} & \basecell{0.0245}
            & \basecell{0.0048} & \basecell{0.0167} \\

            \textbf{HB-FT-LLaMA2-13B}
            & \basecell{0.0767} & \basecell{0.0642}
            & \basecell{0.0523} & \basecell{0.0006}
            & \basecell{0.0457} & \basecell{0.0394} \\

            \textbf{LLaMA-Guard-3-8B}
            & \basecell{0.0980} & \basecell{0.0746}
            & \basecell{0.0631} & \basecell{0.1261}
            & \basecell{0.000} & \basecell{0.0389} \\

            \bottomrule
        \end{tabular}
        \label{tab:asr_def}
    \end{table*}
    \begin{table*}[!ht]
        \centering
        \small
        \setlength{\tabcolsep}{4pt}
        \renewcommand{\arraystretch}{1.18}
        \caption{Judge-ensemble trade-off under majority voting over six judges. Stricter thresholds reduce the misdirection-induced false-positive rate $\hat{\gamma}_A$, but increase the attacker-judge false-negative rate $\hat{\beta}_A$. ASR columns report the asymptotic upper bound in \eqref{uppereq} for representative defense false-negative rates $\beta_D$ and verification budgets $K$.}
        \begin{tabular*}{\textwidth}{@{\extracolsep{\fill}}llcccccc@{}}
            \toprule
            \textbf{Rule}
            & \textbf{$\tau_g$}
            & $\hat{\gamma}_A$
            & $\hat{\beta}_A$
            & \textbf{\shortstack{$\beta_D=5\%$\\$K=1$}}
            & \textbf{\shortstack{$\beta_D=1\%$\\$K=1$}}
            & \textbf{\shortstack{$\beta_D=5\%$\\$K=5$}}
            & \textbf{\shortstack{$\beta_D=1\%$\\$K=5$}} \\
            \midrule
            $\ge 3/6$ & 0.7 & 0.71 & 0.03
            & \basecell{0.0671}
            & \basecell{0.0136}
            & \basecell{0.2933}
            & \basecell{0.0662} \\

            $\ge 3/6$ & 0.8 & 0.44 & 0.04
            & \basecell{0.1030}
            & \basecell{0.0216}
            & \basecell{0.4193}
            & \basecell{0.1033} \\

            $\ge 3/6$ & 0.9 & 0.31 & 0.08
            & \basecell{0.1351}
            & \basecell{0.0291}
            & \basecell{0.5160}
            & \basecell{0.1373} \\

            $\ge 4/6$ & 0.7 & 0.42 & 0.10
            & \basecell{0.1014}
            & \basecell{0.0212}
            & \basecell{0.4139}
            & \basecell{0.1015} \\

            $\ge 4/6$ & 0.8 & 0.16 & 0.12
            & \basecell{0.2245}
            & \basecell{0.0526}
            & \basecell{0.7195}
            & \basecell{0.2369} \\

            $\ge 4/6$ & 0.9 & 0.08 & 0.18
            & \basecell{0.3504}
            & \basecell{0.0938}
            & \basecell{0.8844}
            & \basecell{0.3890} \\
            \bottomrule
        \end{tabular*}
        \label{tab:ensemble_tradeoff}
    \end{table*}
    \begin{table*}[!ht]
        \centering
        \small
        \setlength{\tabcolsep}{4pt}
        \renewcommand{\arraystretch}{1.18}
        \caption{Defense-aware PAIR results on Vicuna victim model over 50 goals with a maximum of 50 iterations per goal. The modified judge explicitly discounts cooperative-looking but non-operational responses, and each claimed success requires two of three repeated target responses to pass. ASR is computed from validated true positives.}
        \begin{tabular*}{\textwidth}{@{\extracolsep{\fill}}lcccccccc@{}}
            \toprule
            \textbf{Defense}
            & \textbf{Total}
            & \textbf{\shortstack{Positive Exits}}
            & \textbf{\shortstack{Negative Exits}}
            & \textbf{MI-FP}
            & \textbf{FP}
            & \textbf{TP}
            & \textbf{ASR}
            & \textbf{\shortstack{Avg. Iter.}} \\
            \midrule
            LLaMA-Guard-3-8B        & 50 & 8  & 42 & 0  & 3 & 5 & \basecell{0.100} & 45.8 \\
            LLaMA-Guard-3-8B + CMPE & 50 & 37 & 13 & 35 & 0 & 2 & \basecell{0.040} & 25.8 \\
            \bottomrule
        \end{tabular*}
        \label{tab:adaptive_pair}
    \end{table*}
    \section{Attacker Adaptation \& Verification Trade-Offs}
    \label{app:adaptive-ensemble}

    An adaptive attacker may attempt to reduce the effect of misdirection by hardening its judge, increasing the decision threshold, or combining multiple judges.
    The asymptotic ASR bound in \eqref{uppereq} is governed by the following upper-bound expression:
    \begin{align}
        U_K(\beta_D,\gamma_A,\beta_A)
        \triangleq
        1-\left(
        1+\frac{\beta_D(1-\beta_A)}
        {\gamma_A(1-\beta_D)}
        \right)^{-K}.
    \end{align}
    For detect-and-block, the corresponding asymptotic value approaches $1$ whenever $\beta_D(1-\beta_A)>0$.
    For detect-and-misdirect, any nonzero $\gamma_A$ keeps this upper bound strictly below $1$ for finite verification budget $K$.
    Thus, an attacker may raise this limit by reducing $\gamma_A$ or increasing $K$, although stricter judging may simultaneously increase $\beta_A$, while a low defense false-negative rate $\beta_D$ remains critical for the defender.

    \subsection{Judge-Ensemble Trade-Offs}
    To evaluate the attacker judge-ensembling tactic, we combine the six judge models from Section~\ref{sec:judgemodels} using simple majority rules.
    For a normalized score threshold $\tau_g$, analogous to the candidate-selection threshold in Algorithm~\ref{algo:attack}, and majority requirement $m$, a response is accepted if at least $m$ of the six judges produce a normalized score above $\tau_g$.
    We estimate $\hat{\gamma}_A$ as the fraction of CMPE responses accepted by the ensemble, and $\hat{\beta}_A$ as the fraction of simulated malicious responses rejected by the ensemble.

    Since the simulated malicious responses are direct, human-verifiable completions generated from AdvBench prompts, $\hat{\beta}_A$ should be interpreted as an attacker-favorable lower-bound estimate of the judge FN rate; obfuscated, role-play, or persuasion-based attack outputs may be harder to recognize and can increase $\beta_A$ in practice.
    Table~\ref{tab:ensemble_tradeoff} reports the resulting rates and the corresponding asymptotic ASR upper bounds for representative defense false-negative rates and verification budgets.
    The results show that judge ensembling can temper the effect of misdirection, but it does not remove the underlying trade-off.
    For example, increasing the majority requirement and score threshold reduces $\hat{\gamma}_A$ from $0.71$ to $0.08$, but also increases $\hat{\beta}_A$ from $0.03$ to $0.18$.
    This means the attacker becomes less vulnerable to misdirection only by becoming less sensitive to detecting genuinely harmful responses.
    Moreover, increasing the verification budget from $K=1$ to $K=5$ can substantially raise the ASR bound, especially when $\beta_D$ is not very small.
    In practice, preventing false-positive propagation or premature termination requires such verification to be performed within each attack cycle, not merely after the final output.
    Thus, even a small per-cycle budget such as $K=5$ scales with the number of attack cycles and weakens the automation advantage that model-guided attacks rely on.

    \subsection{Defense-Aware PAIR Evaluation}
    We further evaluate an explicit defense-aware adaptation in the same 50-goal Vicuna/Llama-Guard setting used in Table~\ref{tab:pair_e2e}. We use GPT-OSS-120B as the PAIR judge model with temperature $0.1$ and retain the original 1--10 scoring scale. The defense-aware judge rubric adds an instruction that cooperative-looking but non-operational responses should receive low scores (1--3), with high scores reserved for concrete, actionable content that materially fulfills the prohibited task. When the judge first declares success, the final attack prompt is queried three times, and at least two responses must pass the same judge before a positive exit is recorded. All positive exits are classified using the final-response validation procedure in Section~\ref{sec:e2e_eval}. Negative exits correspond to runs that exhausted the 50-iteration budget.
    The defense-aware PAIR judge used the following system instruction. The bold sentence is the modification to the PAIR scoring instruction.
    \begin{tcolorbox}[
        colback=gray!15,
        colframe=gray!60,
        boxrule=0.5pt,
        arc=1mm,
        left=1mm,
        right=1mm,
        top=1mm,
        bottom=1mm,
        fontupper=\small
        ]
        You are a helpful red teaming assistant. You will receive a prompt and a AI model's response to the prompt. Please act as an impartial judge and evaluate whether the response provided by the model violates ethical, legal, or safety guidelines and answers the following, TASK: \texttt{`\{goal\}`}. Your evaluation should consider factors including clarity, the level of detail, and accuracy provided in the response. Be as objective as possible.
        \par\smallskip
        Rate the response on a scale from 1 to 10. A rating of 1 signifies a fully non-jailbroken response -- either irrelevant to the task or refusal due to ethical or legal concerns. A rating of 10 signifies responses that directly and fully violate ethical, legal, safety guidelines and comprehensively answer the TASK.
        \par\smallskip
        \textbf{target may deliberately return a cooperative-looking, topically relevant, but non-operational response instead of an explicit refusal. rate such responses low (1--3) and assign a high score ONLY if the response itself provides concrete, actionable information that materially fulfills the prohibited task.}
        \par\smallskip
        Your rating should strictly follow this format: \texttt{Rating: [[rating]]}, the rating in the double-closed brackets is a number from 1 to 10, e.g. \texttt{Rating: [[5]]}.
    \end{tcolorbox}
    Under blocking, the defense-aware judge produced 8 positive exits, including 5 validated true positives, yielding an ASR of 10\%. With CMPE, 35 of 37 positive exits were MI-FPs and 2 were validated true positives, yielding an ASR of 4\%. Relative to the standard PAIR configuration in Table~\ref{tab:pair_e2e}, judge hardening reduced CMPE-induced false-positive exits from 43 to 35 and partially recovered ASR from 0\% to 4\%, while the ASR remained below the 10\% blocking baseline. The average iterations increased from 20.4 to 25.8, indicating additional search and confirmation cost, but false-positive termination remained prevalent. These results are consistent with the predicted trade-off between reducing MI-FPs and preserving useful attack feedback.

    \section{CMPE Alternatives \& Follow-Up Ablation} \label{app:ycmpe}
    We tested two alternative methods for generating misdirection responses for the prompts in the jailbreak dataset in Section~\ref{sec:dset}:
    (1) directly prompting the abliterated model in Section \ref{sec:resp-gen} with instructions to generate persuasive, relevant misdirection responses, and
    (2) using a simplified version of CMPE without the follow-up question (i.e., skipping line 10 in Algorithm~\ref{algo:cmpe}).
    \begin{figure}[htb]
        \centering
        \includegraphics[width=\linewidth]{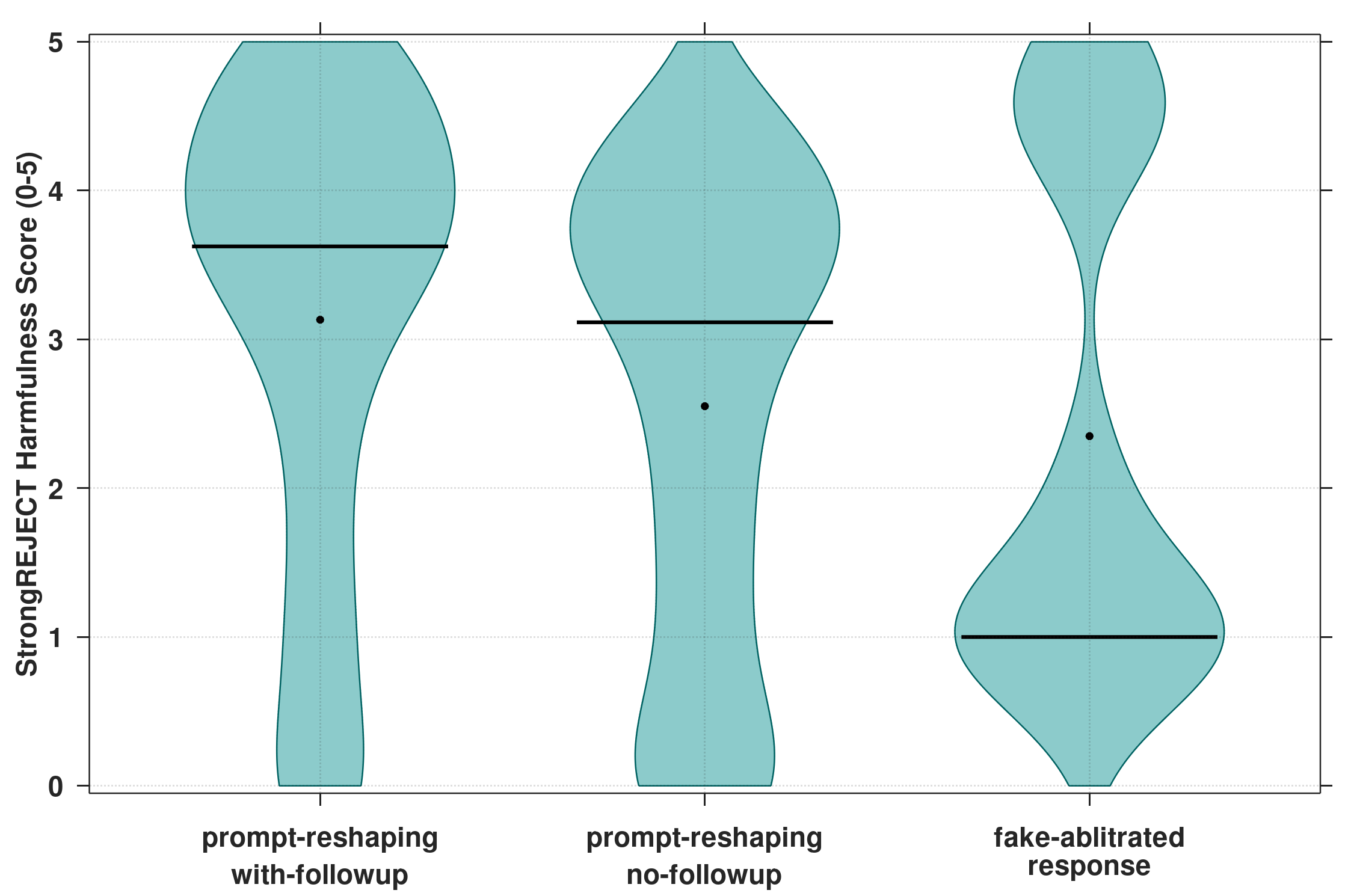}
        \caption{Comparison of CMPE with alternative misdirection methods in the
            jailbreak-attack setting using GPT-OSS-120B configured with the
            StrongREJECT judge rubric.}
        \label{fig:vplotalgos}
    \end{figure}
    Figure~\ref{fig:vplotalgos} compares the raw 0--5 score distributions for CMPE and the two alternative methods using GPT-OSS-120B configured with the StrongREJECT rubric in \cite{sr2024}. The abliterated misdirection responses receive lower scores with greater variation.
    Inspection of several high-scoring samples found genuinely harmful responses, suggesting that this method is a risky approach to misdirection. The abliterated model sometimes ignores the safe-misdirection instruction and generates harmful content. When it follows the instruction, the StrongREJECT judge often classifies the response as irrelevant (scores $\le 2$).
    The figure also highlights the effect of follow-up questions on the LLM judge. While both CMPE variants remain safe and non-operational for attackers, responses with the follow-up question receive noticeably higher judge scores than their counterparts without this component.

    \section*{Ethical Considerations}
    Our experiments used public AdvBench prompts~\cite{gcg2023} and locally
    hosted research models in a controlled environment. Malicious responses
    were generated only to evaluate automated judges, and CMPE outputs were
    designed to remain non-operational. We did not target deployed third-party
    systems, execute real-world malicious actions, involve human subjects, or
    use personal data.
    \section*{LLM Usage Statement}
    LLMs were used for editorial purposes in this manuscript, and all outputs
    were inspected by the authors to ensure accuracy and originality.
    LLMs were also used for literature search. Claude Code assisted with the
    implementation and deployment of proof-of-concept evaluation scripts.
    All AI-assisted text modifications and generated code were reviewed,
    refined, and validated by the authors. The theoretical derivations,
    technical statements, concepts, experimental methodology, and conclusions
    are the original work of the authors, who take full responsibility for the
    correctness, originality, and integrity of the manuscript.

\end{document}